\documentclass[a4paper,fleqn,usenatbib,useAMS]{mnras} 
\usepackage{float}
\usepackage{graphicx}	
\usepackage{amssymb}

\title[A deep \textit{Chandra} observation of the galaxy Arp\,299]{A deep \textit{Chandra} observation of the interacting star-forming galaxy Arp\,299}
\author[K. Anastasopoulou et al.]{
K. Anastasopoulou,$^{1,2}$\thanks{E-mail: kanast@physics.uoc.gr} A. Zezas,$^{1,2,3}$ L. Ballo$^{4}$ and R. Della Ceca$^{4}$\\
$^{1}$Physics Department \& Institute of Theoretical \& Computational Physics, University of Crete, 71003 Heraklion, Crete, Greece\\
$^{2}$Foundation for Research and Technology-Hellas, 71110 Heraklion, Crete, Greece \\
$^{3}$Harvard-Smithsonian Center for Astrophysics, 60 Garden Street, Cambridge, MA 02138, USA\\
$^{4}$Osservatorio Astronomico di Brera (INAF), via Brera 28, I-20121 Milano, Italy}

\date{Accepted XXX. Received YYY; in original form ZZZ}

\pubyear{2016}

\begin{document}
\label{firstpage}
\pagerange{\pageref{firstpage}--\pageref{lastpage}}
\maketitle

\begin{abstract}
We present results from a $\mathrm{90\,ks}$ \textit{Chandra} ACIS-S observation of the X-ray luminous interacting galaxy system Arp\,299 (NGC 3690/IC 694). We detect 25 discrete X-ray sources with luminosities above $\mathrm{\sim 4.0\times 10^{38}\ erg\ s^{-1}}$ covering the entire Ultra Luminous X-ray source (ULX) regime. Based on the hard X-ray spectra of the non-nuclear discrete sources identified in Arp\,299, and their association with young, actively star-forming region of Arp\,299 we identify them as HMXBs.
We find in total 20 off-nuclear sources with luminosities above the ULX limit, 14 of which are point-like sources. Furthermore we observe a marginally significant deficit in the number of ULXs, with respect to the number expected from scaling relations of X-ray binaries with the star formation rate (SFR). Although the high metalicity of the galaxy could result in lower ULX numbers, the good agreement between the observed total X-ray luminosity of ULXs, and that expected from the relevant scaling relation indicates that this deficit could be the result of confusion effects. The integrated spectrum of the galaxy shows the presence of a hot gaseous component with $\mathrm{kT=0.72\pm 0.03\ keV}$, contributing  $\sim$20\% of the soft ($\mathrm{0.1-2.0\ keV}$) unabsorbed luminosity of the galaxy. A plume of soft X-ray emission in the west of the galaxy indicates a large scale outflow. We find that the AGN in NGC 3690 contributes only 22\% of the observed broad-band X-ray luminosity of Arp\,299. 
\end{abstract}


\section{Introduction}

X-ray emission of galaxies is key for understanding compact object populations and accreting binaries and in particular their most extreme manifestation the so-called Ultra-Luminous X-ray sources (ULXs; defined as off-nuclear point sources that have luminosities $\mathrm{L_{(0.1-10.0\ keV)}>10^{39}\ erg\ s^{-1}}$). In particular, high star-formation rate galaxies often host large populations  of ULXs \citep[e.g.][]{swartz11}. As such they are useful for studying the population, demographics, and scaling relations of ULXs with galaxy parameters. Since scaling relations are based on local galaxies with star formation rates (SFR) up to a few ($\sim{20}$) $\mathrm{M_{\odot}/yr}$ \citep[e.g.][]{m1,leh10}, and there is some evidence for evolution of these scaling relations with SFR \citep{antara,m3}, examining if a local highly star-forming galaxy verifies these relations is an important test of their applicability to the most extreme star forming galaxies and is informative for their use as analogues of high-z galaxies.

Arp\,299 is one of the most powerful star-forming galaxies in the local Universe  \citep[44Mpc;][]{b3} belonging to the class of Luminous Infrared Galaxies \citep[LIRGs;][]{b4} with a total infrared luminosity of $\mathrm{L_{IR}=5.16\times 10^{11}L_{\odot}}$ \citep{b6}. It consists of two galaxies in an advanced merging state separated by $\sim$22". The western galaxy is referred to as NGC 3690 (or B) and it is believed to host an AGN \citep{b9,z03,ballo,ptak}. The eastern galaxy is referred to as IC 694 (or A) and the overlapping region as C and C' \citep[following the nomenclature of ][]{hy}.

Due to its proximity and intense star forming activity Arp\,299 has been studied extensively in all wavelengths. Optical and infrared observations have shown that Arp\,299 is dominated by widespread star formation taking place in the two nuclei and in the overlapping area with typical ages of $\sim$ 15Myrs \citep{ah00}. Results in the mid infrared show that most of the star-forming regions are deeply embedded into dust \citep{b6}. Recent \textit{Spitzer}/IRS results are now showing that  the integrated mid-infrared spectrum of Arp\,299 exhibits a remarkable similarity with those of high-z Ultra Luminous Infrared Galaxies (ULIRGs) albeit its lower luminosity and possibly higher metallicity. This suggests that it may represent a local example of the star-forming processes occurring at high-z \citep{ah09}.

In the X-ray regime Arp\,299 is the second most luminous galaxy in the local Universe ($\lesssim 50\ \mathrm{Mpc}$) with a luminosity of $\mathrm{L(0.1-10.0\ keV)=4\times 10^{41}ergs\ s^{-1}}$ \citep{z98,b3}. A short $\mathrm{24\,ks}$ \textit{Chandra} observation obtained in 2001 showed that Arp\,299 hosts 16 ULXs, one of the richest galaxies in the local Universe, although the short exposure allowed to probe only a fraction of the overall population \citep{z03}.
\citet{luan} studied a sample of 17 nearby LIRGs (including Arp\,299) and found a general deficit in their number of ULXs per unit SFR compared to the rate in nearby normal star-forming galaxies from \citet{swartz11}. This result is also supported by \citet{smith} who found that the total number of ULXs in LIRGs (including Arp\,299) normalised to their far-infrared luminosity is deficient in comparison to the ULX rates found for spiral galaxies. They argue that metallicity may have some influence on ULX numbers but the main reason for this deficit is high columns of gas and dust that obscure these ULXs from our view. If this trend holds for galaxies experiencing intense starbursts, it will have important implication for understanding the X-ray output of high-z galaxies.

Furthermore \textit{BeppoSAX} revealed for the first time the existence of a deeply buried ($\mathrm{N_{H}\sim 2.5\times 10^{24} cm^{-2}}$) AGN with a luminosity of $\mathrm{L_{0.5-100\ keV}\simeq 1.9\times 10^{43} erg\ s^{-1}}$ \citep{b9}. \textit{Chandra} and \textit{XMM-Newton} observations confirmed the existence of a strongly absorbed AGN and located it in the nucleus of NGC 3690 while there is evidence that the second nucleus IC 694 might also host an AGN of lower luminosity \citep{z03, ballo, b11, perez, ah13}. Recent results of simultaneous observations with \textit{NuSTAR} and \textit{Chandra} (including \textit{Chandra} data presented in this paper) confirmed the existence of an AGN in NGC 3690 constrained its total X-ray luminosity to $\mathrm{L(10-30\ keV)\sim 1.2 \times 10^{43} erg s^{-1}}$ and its obscuring column density to $\mathrm{N_H\sim4 \times 10^{24} cm^{-2}}$. It also showed that any AGN in IC 694 must be heavily obscured or have a much lower luminosity than that in NGC 3690 \citep{ptak}. 
  
In this paper we analyse data from a deep \textit{Chandra} observation of Arp\,299 ($\simeq 90\,\mathrm{ks}$) which allow us to further explore the nature of the X-ray source population in this interesting system. In particular we are able to study the entire population of ULXs by reaching a detection limit of $\mathrm{4\times 10^{38}\ erg\ s^{-1}}$ and to explore the contribution of the AGN to the energetics of the galaxy. Arp\,299 is possibly an analogue of actively star-forming galaxies at higher redshifts \citep{ah09}. Therefore studying its hot gaseous and X-ray binary components and energetics with this deep \textit{Chandra} high resolution observation and their scaling with its star-forming activity we can have useful insights of the nature of the high-z normal galaxies detected in medium and high depth surveys where a detailed analysis of the sources is not possible.

The structure of the paper is as follows: In section \ref{observationanddataanalysis} we describe the observation, the data analysis and our results. We discuss our results in section \ref{discussion} and in section \ref{summary} we summarize our findings. All errors from spectral analysis correspond to the 90\% confidence interval unless otherwise stated.

\section{Observation and Data Analysis}\label{observationanddataanalysis}

\textit{Chandra} \citep{weisskopf} observed  Arp\,299 with the ACIS-S camera \citep{b1} for a total of 90.37$\,\textrm{ks}$. The observation was split in two segments due to scheduling constraints: a 38.49$\,\textrm{ks}$ exposure performed on the 12th of March 2013 (OBSID 15619) and a second exposure of 51.88$\,\textrm{ks}$ performed on the 13th of March 2013 (OBSID 15077). 
The data analysis was performed with the CIAO software version 4.7 and CALDB version 4.6.5. To apply the latest calibration data we reprocessed the Level-1 events files using the \textit{acis\_process\_events} tool. We then created the Level-2 events files using the \textit{dmcopy} tool to filter for bad grades and status bits (keeping only grades=0, 2, 3, 4, 6 and status=0). We checked for background flares during our observation and we found that the background level is fairly constant.

Since the two observations are almost contiguous, and they are performed at the same ROLL angle and pointing, we created an events file for the merged exposure. Using the \textit{reprojec\_obsid} tool we first reprojected the two Level-2 files to a common tangent point and then we combined them to create the merged Level-2 events file. The scientific analysis was performed on these three files.

We created images as well as exposure maps in the broad (0.5-7.0 \textrm{keV}), soft (0.5-1.2 \textrm{keV}), medium (1.2-2.0 \textrm{keV}), and hard (2.0-7.0 \textrm{keV}) bands using the \textit{fluximage} tool for the reprojected files and the \textit{flux\_obs} tool for the merged one. We additionally normalized all the exposure maps to the exposure of a reference pixel at approximately the centre of the galaxy.

\begin{figure*}
	\centering
	\begin{minipage}{140mm}
	\resizebox{\hsize}{!}{\includegraphics[scale=1]{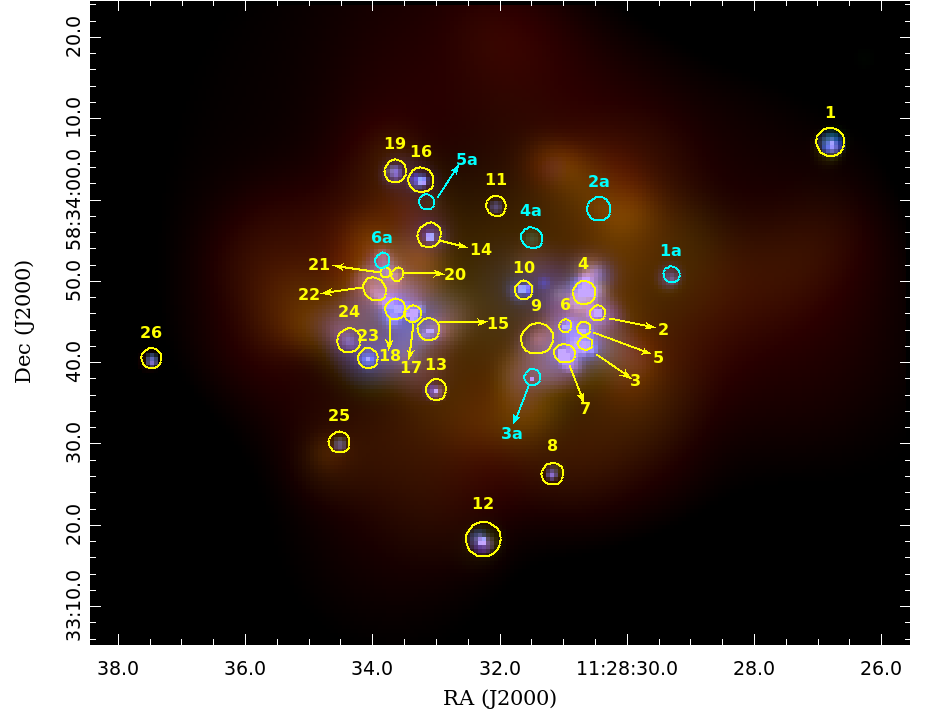}}
	 \caption{An adaptively smoothed true colour image of Arp\,299 from the merged observation. We show with red colour the soft band (0.5-1.2 \textrm{keV}), with green colour the medium band (1.2-2.0 \textrm{keV}), and with blue the hard band (2.0-7.0 \textrm{keV}). Overlaid are the 26 sources with $\mathrm{SNR>3.0}$ (yellow circles) as well as the 6 detections (1a to 6a; cyan circles) with $2.0<$SNR$<3.0$ that are noted as possible sources.}
	\label{fig.smoothcolor}
\end{minipage}
\end{figure*}

We adaptively smoothed these images using the \textit{csmooth} CIAO tool with a Gaussian convolution kernel, and applying a minimum and maximum signal-to-noise ratio of 3 and 5 respectively. A ``true colour image'' consisting of the soft (red), medium (green), and hard (blue) smoothed images for the merged observation is shown in Fig. \ref{fig.smoothcolor}. From this image we see a population of discrete sources (\S \ref{sourcedetection}) and extended soft diffuse emission of the galaxy. 

\subsection{Source Detection}\label{sourcedetection}
We used the \textit{wavdetect} tool \citep{b2} to detect discrete sources. We searched in the broad (0.5-7.0 \textrm{keV}), soft (0.5-1.2 \textrm{keV}), medium (1.2-2.0 \textrm{keV}), and hard band (2.0-7.0 \textrm{keV}) images of the reprojected individual observation segments, as well as, the merged dataset for sources on scales of 2, 4, 8, and 16 pixels. We then cross-correlated the source lists created in all bands of the individual and the merged observations. We visually inspected the final source list in order to make sure that no individual sources were excluded.
We found in total 42 sources within the D25 region of the galaxy \citep[RC3;][]{rc31,rc32}.

In order to calculate the photometric parameters of the sources, we manually defined a source aperture based on the merged observation. We took care to include as many of the source counts as possible and at the same time to avoid contamination from nearby sources and background. The minimum aperture radius of ($\sim 1''$) which encompasses at least 90\% of the  encircled energy at $\mathrm{1.49\ keV}$ of an on-axis point source for the ACIS-S camera was satisfied by the mean aperture radius of our sources which is $\sim 1.3''$. The background was defined as an annulus around each source with inner radius about 1-2 pixel larger than the source aperture in order to avoid contamination by the wings of the source PSF and outer radius large enough ($\sim$ 2-5$''$) to obtain good count statistics to perform photometry and to avoid contamination from nearby sources.

\begin{table*}
	\centering
		\begin{minipage}{140mm}
		\caption{Properties of the discrete sources in the broad band ($\mathrm{0.5-7.0\ keV}$).}
		\begin{tabular}{@{}cccrrccccc@{}}
			\hline   
			Src\footnote{Column 1: The source identification number, Columns 2 and 3: Sky coordinates, Column 4: Net source counts and corresponding error counts, Columns 5 and 6: The background source counts and the signal to noise ratio of each source, Columns 7 and 8: The two ellipse major and minor radius for the source apertures, Column 9: The corresponding source identifier from \citet{z03}, Column 10: Comments on variability, extension, and name sources from previous works. Sources that are point-like ULXs are also reported in this column.}  &  RA      & Dec   &       Net counts& Bkg & S/N & $\mathrm{r_1}$  &   $\mathrm{r_2}$  & Src ID& Notes \\
			ID &h m s &$^{\circ}$ $'$ $''$   & $\pm$error & & &$''$&$''$ & (Z03)&\\
			(1) &  (2)      & (3)   &       (4)& (5) & (6) &(7)  &   (8)  & (9)& (10)\\
			\hline
			1   & 11:28:26.8  & +58:34:07 &135.3  $\pm$ 12.9   & 4.7   & 10.1  & 1.74 & 1.73 &    1& ULX        \\
			2   & 11:28:30.4  & +58:33:46 &430.8  $\pm$ 25.3   & 115.2 & 15.9  & 0.93 & 0.94 &    -&     	ULX   \\
			3   & 11:28:30.6  & +58:33:42 &336.2  $\pm$ 28.7   & 260.7 & 10.9  & 0.92 & 0.77 &   4? &  ext (hard)      \\
			4   & 11:28:30.7  & +58:33:48 &583.0  $\pm$ 37.8   & 372.0 & 15.4  & 1.43 & 1.39 &    2&   C; ext (hard)     \\
			5   & 11:28:30.7  & +58:33:44 &206.2  $\pm$ 19.8   & 84.8  & 9.9   & 0.82 & 0.82 &   3 &   ext (hard)     \\
			6   & 11:28:31.0  & +58:33:44 &140.5  $\pm$ 16.7   & 80.5  & 7.5   & 0.73 & 0.80 &   5 &    ULX    \\
			7   & 11:28:31.0  & +58:33:41 &892.5  $\pm$ 37.3   & 264.4 & 22.8  & 1.31 & 1.11 &   6 &AGN; B1; ext     \\
			8   & 11:28:31.2  & +58:33:26 &35.2   $\pm$ 7.5    & 5.8   & 4.2   & 1.33 & 1.36 &   7 &        \\
			9   & 11:28:31.4  & +58:33:43 &72.2   $\pm$ 24.5   & 172.7 & 3.3   & 2.00 & 1.85 &    - &  ULX  \\
			10  & 11:28:31.6  & +58:33:49 &221.3  $\pm$ 16.9   & 23.6  & 12.5  & 1.08 & 1.13 &    8&   ULX; C''     \\
			11  & 11:28:32.0  & +58:33:59 &23.2   $\pm$ 6.8    & 6.8   & 3.0   & 1.27 & 1.18 &    - &        \\
			12  & 11:28:32.2  & +58:33:18 &198.2  $\pm$ 15.6   & 12.8  &  12.2  & 2.11 & 2.15 &  9 &      ULX   \\
			13  & 11:28:33.0  & +58:33:37 &70.3   $\pm$ 11.1   & 24.6  & 5.7   & 1.23 & 1.30 &   10 &     ULX   \\
			14  & 11:28:33.1  & +58:33:56 &195.0  $\pm$ 16.6   & 30.0  & 11.2  & 1.40 & 1.54 &    11&   ULX \\
			15  & 11:28:33.1  & +58:33:44 &147.6  $\pm$ 19.3   & 99.4  & 7.4   & 1.36 & 1.34 &    12&    ULX    \\
			16  & 11:28:33.2  & +58:34:02 &213.7  $\pm$ 16.4   & 8.2   & 13.0  & 1.57 & 1.47 &    13&    ULX    \\
			17  & 11:28:33.3  & +58:33:45 &559.9  $\pm$ 28.1   & 109.0 & 19.1  & 1.04 & 0.99&    14&     ULX   \\
			18  & 11:28:33.6  & +58:33:46 &423.3  $\pm$ 36.9   & 294.6 & 12.7  & 1.28 & 1.20 &    16& A; ext (hard)    \\
			19  & 11:28:33.6  & +58:34:03 &56.7   $\pm$ 10.3   & 17.3  & 5.1   & 1.31 & 1.41 &    -& ULX;variable\footnote{See \S \ref{fig.variable}.}       \\
			20  & 11:28:33.6  & +58:33:51 &43.3   $\pm$ 12.1   & 58.7  & 3.0   & 0.85 & 0.71 &    15& ext        \\
			21  & 11:28:33.8  & +58:33:51 &45.3   $\pm$ 11.5   & 40.6  & 3.6   & 0.67 &  -   &    15&        \\   
			22  & 11:28:34.0  & +58:33:49 &132.1  $\pm$ 25.4   & 238.8 & 5.0   & 1.54 & 1.27 &    -& blob; ext (hard)   \\
			23  & 11:28:34.1  & +58:33:40 &133.1  $\pm$ 15.0   & 45.8  & 8.1   & 1.21 & 1.23 &   17 &   ULX     \\
			24  & 11:28:34.4  & +58:33:43 &50.0   $\pm$ 12.8   & 61.9  & 3.4   & 1.40 & 1.51 &   - &   ext (hard)     \\
			25  & 11:28:34.5  & +58:33:30 &26.5   $\pm$ 7.0    & 7.4   & 3.3   & 1.29 & 1.28 &    -&        \\
			26  & 11:28:37.5  & +58:33:40 &35.6   $\pm$ 7.2    & 1.4   & 4.7   & 1.21 & 1.23 &   18&   ULX     \\
			\hline			
			\end{tabular} 	
		\label{tab.properties}
	\end{minipage}
	\end{table*}

 We used the \textit{dmextract} tool to perform the photometry on the 42 sources in the merged and individual exposures for the broad, hard, medium, and soft bands. In more detail we ran  \textit{dmextract} on the images and normalized exposure maps for the discrete sources and on the background image. The errors on the number of counts were calculated following the Gehrels approximation \citep{gehrels}.
We calculated the signal to noise ratio (SNR) of each source as $\mathrm{SNR=S/\sqrt{(1+\sqrt{0.75+T})^2+(1+\sqrt{0.75+B})^2}}$ where S are the net source counts, T the total counts, and B the estimated background counts for each source rescaled to the source area. We consider as significant detections and therefore present in this paper the results for the 26 sources that have $\textrm{SNR}\geq 3.0$. We also present six detections (1a-6a) that exceeded $\mathrm{SNR=2.0}$ as possible sources. 

The photometry for the 26 sources obtained from the merged events file is presented in  Table \ref{tab.properties} for the broad band. 
Column (1) gives the source number, Columns (2) and (3) give the sky coordinates, Column (4) gives the net source counts and the corresponding error counts. The background source counts  and the signal to noise ratio of each source  are presented in columns (5) and (6). Columns (7) and (8) give the ellipse major and minor radius for the source apertures. Column (9) gives the corresponding source identifier from \citet{z03}, and Column (10) presents comments about each source such as variability, extension, and names from previous works. The same information for the hard, medium, and soft bands is presented in the Appendix (Table \ref{tab.properties2}). The description of columns is the same as for columns (1)-(6) of Table \ref{tab.properties}. Columns (7)-(10) of Table \ref{tab.properties} are exactly the same and thus not included. The properties of the 6 lower significance detections (1a to 6a), that are noted as possible sources, are presented in the Appendix (Table \ref{tab.properties1a6a}).

The location of 26 discrete sources as well as the 6 lower significance detections (1a to 6a) is shown on the ``true colour image'' of Fig. \ref{fig.smoothcolor}. 
\begin{figure}
	\resizebox{\hsize}{!}{\includegraphics[scale=1]{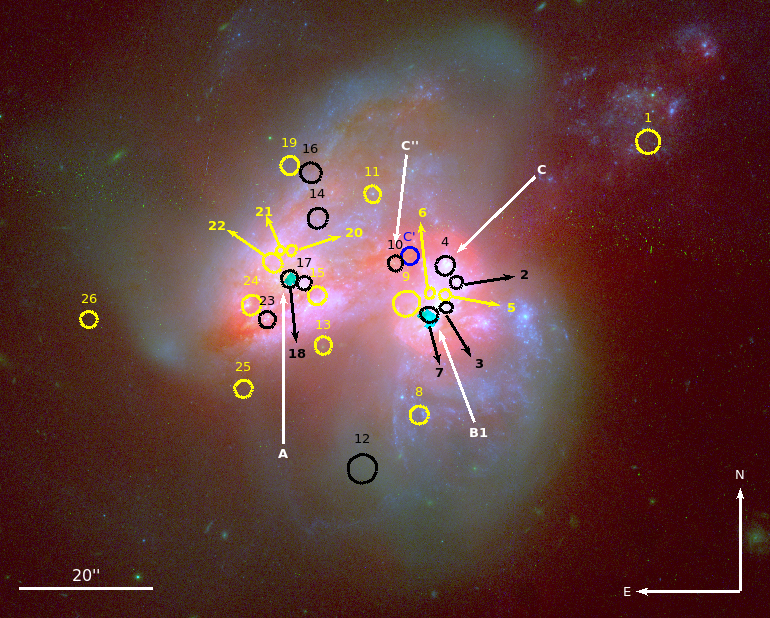}}
	\caption{Colour image of Arp\,299. Composed of a F814W ($\mathrm{814\ nm}$; green), F435W ($\mathrm{435\ nm}$; blue) HST images and IRAC band 4 ($\mathrm{8\ \mu m}$; red; saturated at the nuclei) non stellar image (see text). The X-ray sources with luminosities $\mathrm{L(0.1-10.0\ keV)>5\times 10^{39}\ erg\ s^{-1}}$ are shown with black circles, while lower luminosity sources ($\mathrm{L(0.1-10.0\ keV)<5\times 10^{39}\ erg\ s^{-1}}$) are shown with yellow circles. With white arrows we show the additional names of our sources Src 18 (A), Src 7 (B1), Src 4 (C), Src 10 (C"); C' is unidentified in our observation and is presented here with a blue circle.} 
	\label{fig.color}
	
\end{figure}
In Fig. \ref{fig.color} we overlay the location of the discrete sources on a multi-band image consisting of HST images in the F814W ($\mathrm{814\ nm}$; green) and F435W ($\mathrm{435\ nm}$; blue) and a \textit{Spitzer} IRAC band 4 ($\mathrm{8\ \mu m}$; red) non-stellar image. The IRAC non stellar image in particular was produced by subtracting a scaled $\mathrm{3.6\ \mu m}$ image from the $\mathrm{8\ \mu m}$ in order to remove the stellar emission \citep{br}. The above filters give a picture of the distribution of the old, young, and obscured young stellar populations respectively (Fig. \ref{fig.color}). We see that the vast majority of the sources are located within the main body of the galaxy and particularly its most actively star-forming regions.

\subsection{Spectral Analysis}\label{spectralanalysis}

We extracted source and background spectra of the discrete sources using the \textit{specextract} tool. Since the two observations are taken with the same orientation, and within 13 hours of each other, we decided to extract the spectra from the merged events file \footnote{\url{http://cxc.harvard.edu/ciao/caveats/merged_events.html}}
. This is supported by the fact that the spectral parameters of the sources are consistent between the two observations, while all but one source do not show any variability between the observations (\S \ref{variability}).
We supplied \textit{specextract} a stack of the two reprojected events files and a stack of their corresponding auxiliary files, and we ran it with \textit{combine=yes}. This produced the correctly summed Pulse Height Amplitude (PHA) file, spatially weighted Auxiliary Response File (ARF) and Redistribution Matrix File (RMF). As a final step we grouped the resulting spectra to have at least 20 total counts per spectral bin using the \textit{dmgroup} tool to allow for $\chi^2$ fitting.

We fitted the grouped spectra using the XSPEC v12.8.2 software \citep{xspec}. We ignored events with energies above $\textrm{8.0 keV}$ and below $\textrm{0.4 keV}$ since they are dominated by the background. The spectral fitting was performed only for the 20 sources with more than about $\approx$50 net counts (i.e 3 or more spectral bins). In all but four cases a single power-law model which is generally used to fit the spectra of X-ray binaries (XRBs) with photoelectric absorption gave satisfactory fits. For these sources the photon indices range from $\Gamma= 0.9$ to $\Gamma=3.9$, consistent with those of XRBs, and the hydrogen column density is typically greater than the Galactic ($\sim 9.5\times 10^{19} \mathrm{cm^{-2}}$; using the Colden tool\footnote{\url{http://cxc.harvard.edu/toolkit/colden.jsp}}).

The spectra of sources 9 and 22 were too noisy after subtracting their local background. 
Since we extract the background from areas surrounding each source, this indicates that these two sources are local enhancements in the diffuse emission and their spectra are similar to the spectra of their surrounding regions. Therefore, in order to obtain a picture of the physical properties of these regions we extracted background from an area outside the galaxy and we model the diffuse emission with a thermal plasma component.

For Src 9 an absorbed power-law plus a thermal plasma model gave a good fit while for Src 22 a reasonable fit was given by an absorbed thermal plasma model. Source 9 could be a diffuse emission region but the existence of a power-law component suggests that one or more X-ray binaries are embedded in this region. We consider Src 22 a diffuse emission region as its spectrum is well fitted with only a thermal plasma component. The detailed results for the fitted parameters of the 20 sources are reported in Table \ref{tab.spectralparam}.

\begin{table*}
	\centering
	\begin{minipage}{140mm}
		\caption{Spectral parameters based on spectra and X-ray colours of the discrete sources.}
		\hskip-2.5cm\begin{tabular}{@{\extracolsep{5pt}}cccccrrrcc@{}}
			\hline 
			&\multicolumn{4}{c}{Spectral analysis}&\multicolumn{5}{c}{X-ray colours}\\
			 \cline{2-5}
			 \cline{6-10}
			  	&&&&&&&&&\\
			Src ID\footnote{Column 1: Source identification number, Column 2: Photon index values from spectral fitting and corresponding errors, Column 3: Thermal plasma model temperatures, Column 4: Line-of-sight hydrogen column density, Column 5: Chi-square of the spectral fit and degrees of freedom (d.o.f), Columns 6, 7, and 8: X-ray colours, and their corresponding uncertainties, defined as $C_1=\log_{10} ({S}/{M})$, $C_2=\log_{10} ({M}/{H})$, $C_3=\log_{10} ({S}/{H})$ where S, M, and H are the net counts in the soft (0.5-1.2 \textrm{keV}), medium (1.2-2.0 \textrm{keV}), and hard (2.0-7.0 \textrm{keV}) bands (see text for details). Columns 9 and 10: Photon index $\mathrm{\Gamma_{col}}$ and hydrogen column density $\mathrm{N_{Hcol}}$ based on the X-ray colours} & $\mathrm{\Gamma}$ & $\mathrm{kT}$  & $\mathrm{N_H}$ & $\chi^2$ (dof) & $\mathrm{C_1}$ &  $\mathrm{C_2}$ & $\mathrm{C_3}$ & $\mathrm{\Gamma_{col}}$  &  $\mathrm{N_{H col}}$ \\
			&          &  keV  &  $\mathrm{10^{22} cm^{-2}}$     & & &   & & &$\mathrm{10^{22} cm^{-2}}$ \\
			(1)&(2)&(3)& (4)& (5)& (6)& (7) &(8)& (9)&(10)\\ 
			\hline			
			1       &              $2.19_{-0.63}^{+0.89}$  &-     &                       $0.86_{-0.45}^{+0.88}$  &                       2.6       (4)    &     $-0.50_{-0.20}^{+0.20}$  &                        $-0.02_{-0.13}^{+0.13}$  &                        $-0.52_{-0.20}^{+0.20}$                   &                        2.0                     &                   0.6                      \\[5pt]
			2       &              $1.86_{-0.32}^{+0.36}$  &-     &                       $0.23_{-0.16}^{+0.19}$  &                       22.3      (21)   &     $-0.12_{-0.11}^{+0.11}$  &                        $-0.00_{-0.09}^{+0.09}$  &                        $-0.12_{-0.11}^{+0.11}$                   &                        1.6                     &                    0.13                     \\[5pt]
			3       &              $3.20_{-0.66}^{+0.78}$  &-     &                       $2.33_{-0.70}^{+0.87}$  &                       29.1      (24)   &     $-1.62_{-1.48}^{+0.92}$  &                        $-0.22_{-0.12}^{+0.12}$  &                        $-1.84_{-1.47}^{+0.91}$                   &                        3.5                     &                    3.0                      \\[5pt]
			4       &              $2.76_{-0.50}^{+0.60}$  &-     &                       $0.27_{-0.15}^{+0.17}$  &                       37.7      (37)   &     $                        0.00_{-0.10}^{+0.10}$    &                        $0.31_{-0.12}^{+0.12}$   &                                         $0.32_{-0.12}^{+0.13}$   &                       2.7                  &                        0.25      \\[5pt]
			5       &              $2.67_{-0.89}^{+1.06}$  &-     &                       $0.88_{-0.68}^{+0.83}$  &                       7.8       (9)    &     $-0.49_{-0.26}^{+0.23}$  &                        $0.12_{-0.14}^{+0.14}$   &                        $-0.36_{-0.27}^{+0.24}$                   &                        2.6                     &                    0.8                      \\[5pt]
			6       &              $3.92_{-1.17}^{+1.69}$  &-     &                       $2.89_{-1.28}^{+1.97}$  &                       7.4       (8)    &     $-1.49_{-1.50}^{+0.97}$  &                        $-0.21_{-0.16}^{+0.17}$  &                        $-1.70_{-1.49}^{+0.95}$                   &                        3.2                     &                    2.5                      \\[5pt]
			7       \footnote{The  spectral                model  contains                also                    a                       Gaussian  line   with  $\mathrm{E_l=6.30\pm     0.03\                    keV}$                    with                     $\mathrm{\sigma=0.018_{-0.018}^{+0.093}\  keV}$                    and                     $\mathrm{EW=0.72\pm  0.01\                    keV}$.}&  $1.02_{-0.75}^{+0.66}$   &        $0.52_{-0.18}^{+0.09}$   &  $1.02_{-0.27}^{+0.22},  127.0_{-61.9}^{+51.9}$\footnote{Equivalent  hydrogen  column   density  of  partial  covering  fraction  absorption  with  covering  fraction  95\%.}  &  52.53  (41)  &  $-0.24_{-0.09}^{+0.09}$  &  $-0.10_{-0.06}^{+0.06}$  &  $-0.34_{-0.08}^{+0.09}$  &  -  &  -  \\[5pt]
			8       &              -                       &      -                       &                       -                       &         -      &     $-0.39_{-0.56}^{+0.46}$  &                        $0.15_{-0.28}^{+0.29}$   &                        $-0.24_{-0.56}^{+0.50}$                   &                        2.6                     &                    0.6                      \\[5pt]
			9       &              $2.73_{-1.72}^{+0.97}$  &      $0.46_{-0.16}^{+0.52}$  &                       $0.51_{-0.27}^{+0.16}$  &         13.9   (7)   &                        $                        0.76_{-0.72}^{+1.15}$    &                        $0.28_{-1.82}^{+1.86}$                    &                        $1.07_{-0.82}^{+1.42}$  &                    -                        &         -                        \\[5pt]
			10      &              $1.21_{-0.35}^{+0.88}$  &-     &                       $0.06_{-0.06}^{+0.91}$  &                       10.8      (7)    &     $-0.77_{-0.28}^{+0.26}$  &                        $-0.16_{-0.11}^{+0.11}$  &                        $-0.92_{-0.28}^{+0.25}$                   &                        1.9                     &                    1.0                      \\[5pt]
			11      &              -                       &      -                       &                       -                       &         -      &     $-0.52_{-1.19}^{+0.75}$  &                        $0.07_{-0.40}^{+0.39}$   &                        $-0.45_{-1.19}^{+0.77}$                   &                        2.5                     &                    0.8                      \\[5pt]
			12      &              $1.91_{-0.67}^{+0.78}$  &      -                       &                       $0.40_{-0.32}^{+0.40}$  &         4.3    (6)   &                        $-0.26_{-0.14}^{+0.14}$  &                        $0.11_{-0.12}^{+0.12}$   &                                         $-0.14_{-0.15}^{+0.15}$  &                       2.2                  &                        0.40      \\[5pt]
			13      &              $1.98_{-0.69}^{+1.02}$  &      -                       &                       $0.14_{-0.14}^{+0.41}$  &         0.5    (2)   &                        $                        0.00_{-0.27}^{+0.27}$    &                        $0.10_{-0.25}^{+0.26}$                    &                        $0.10_{-0.28}^{+0.28}$  &                    1.7                      &         0.07                     \\[5pt]
			14      &              $1.72_{-0.73}^{+0.84}$  &-     &                       $0.53_{-0.48}^{+0.63}$  &                       3.0       (6)    &     $-0.36_{-0.22}^{+0.20}$  &                        $-0.07_{-0.12}^{+0.12}$  &                        $-0.44_{-0.22}^{+0.20}$                   &                        1.6                     &                    0.4                      \\[5pt]
			15      &              $2.67_{-1.32}^{+2.01}$  &      -                       &                       $0.96_{-0.82}^{+1.36}$  &         11.9   (8)   &                        $-0.57_{-0.60}^{+0.42}$  &                        $0.18_{-0.18}^{+0.18}$   &                                         $-0.40_{-0.62}^{+0.44}$  &                       3.0                  &                        1.0       \\[5pt]
			16      &              $2.53_{-0.74}^{+0.85}$  &-     &                       $1.28_{-0.53}^{+0.68}$  &                       6.2       (6)    &     $-0.55_{-0.21}^{+0.20}$  &                        $-0.08_{-0.10}^{+0.11}$  &                        $-0.62_{-0.21}^{+0.20}$                   &                        1.8                     &                    0.6                      \\[5pt]
			17      &              $1.69_{-0.28}^{+0.31}$  &-     &                       $0.25_{-0.14}^{+0.16}$  &                       27.7      (28)   &     $-0.25_{-0.09}^{+0.09}$  &                        $0.01_{-0.08}^{+0.08}$   &                        $-0.24_{-0.09}^{+0.09}$                   &                        1.8                     &                    0.3                      \\[5pt]
			18      &              $0.93_{-0.60}^{+0.70}$  &-     &                       $1.87_{-1.00}^{+1.47}$  &                       40.10     (31)   &     $-1.46_{-1.55}^{+1.01}$  &                        $-0.67_{-0.20}^{+0.19}$  &                        $-2.13_{-1.50}^{+0.98}$                   &                        -                       &                    -                        \\[5pt]
			19      &              $1.02_{-0.47}^{+0.92}$  &-     &                       $0.01_{-0.01}^{+0.42}$  &                       1.8       (1)    &     $-0.43_{-0.88}^{+0.56}$  &                        $-0.12_{-0.24}^{+0.23}$  &                        $-0.55_{-0.87}^{+0.54}$                   &                        1.5                     &                    0.45                     \\[5pt]
			20      &              -                       &      -                       &                       -                       &         -      &     $                        0.33_{-0.68}^{+0.84}$    &                        $0.19_{-0.86}^{+0.68}$   &                                         $0.52_{-0.54}^{+0.54}$   &                       -                    &                        -         \\[5pt]
			21      &              -                       &      -                       &                       -                       &         -      &     $-1.09_{-1.45}^{+0.88}$  &                        $0.46_{-0.35}^{+0.40}$   &                        $-0.61_{-1.61}^{+1.01}$                   &                        -                       &                    -                        \\
			22      &              -                       &      $0.63_{-0.07}^{+0.17}$  &                       $0.62_{-0.08}^{+0.08}$  &         33.6   (16)  &                        $                        0.47_{-0.34}^{+0.41}$    &                        $1.30_{-1.11}^{+1.72}$                    &                        $1.79_{-0.99}^{+1.53}$  &                    -                        &         -                        \\[5pt]
			23      &              $2.02_{-0.54}^{+0.53}$  &      -                       &                       2.05                    &         2.9    (5)   &                        $-0.84_{-1.13}^{+0.63}$  &                        $-0.36_{-0.18}^{+0.17}$  &                                         $-1.21_{-1.11}^{+0.60}$  &                       1.3                  &                        1.0       \\[5pt]
			24      &              $2.69_{-1.91}^{+3.89}$  &-     &                       0.0095\footnote{This    value                   is        fixed  at    the                      Galactic                 line-of-sight            value.}                  &                                         1.1                      (4)                     &                    $-0.19_{-1.81}^{+1.52}$  &         $-0.62_{-1.02}^{+0.58}$  &        $-0.83_{-1.34}^{+0.74}$  &  -                       &                                           -         \\[5pt]
			25      &              -                       &      -                       &                       -                       &         -      &     $-0.12_{-0.47}^{+0.41}$  &                        $0.37_{-0.44}^{+0.46}$   &                        $0.24_{-0.56}^{+0.58}$                    &                        3.2                     &                    0.45                     \\[5pt]
			26      &              -                       &      -                       &                       -                       &         -      &     $-0.16_{-0.32}^{+0.32}$  &                        $-0.05_{-0.28}^{+0.28}$  &                        $-0.21_{-0.32}^{+0.31}$                   &                        1.4                     &                    0.13                     \\[5pt]
			\hline
			\end{tabular} 	
		\label{tab.spectralparam}
	\end{minipage}
\end{table*}

\subsection{Hardness ratios}\label{hardnessratios}

Hardness ratios or X-ray colours are a useful tool for deriving the spectral properties of faint sources. 
We calculated X-ray colours for all sources; in this way we estimate the spectral parameters for the sources with less than 50 counts while for the brighter sources we can have a direct comparison of the spectral parameters calculated from the X-ray colours and the spectral fits. X-ray colours are defined as $\mathrm{C_1\equiv\log_{10} ({S}/{M})}$, $\mathrm{C_2\equiv\log_{10} ({M}/{H})}$, $\mathrm{C_3\equiv\log_{10} ({S}/{H})}$ where S, M, and H  are the net counts in the soft (0.5-1.2 \textrm{keV}), medium (1.2-2.0 \textrm{keV}), and hard (2.0-7.0 \textrm{keV}) bands. In our analysis we use X-ray colours instead of the hardness ratios, since they are better behaved in terms of their error distribution \citep{park}.
We calculated the X-ray colours and their uncertainties for the 26 sources, using the BEHR code \citep{park} which evaluates their posterior probability distribution and provides reliable estimates and confidence limits even
when either or both soft and hard counts are very low. The resulting X-ray colours and their corresponding  90\% confidence intervals, are presented in Table \ref{tab.spectralparam}.

\begin{figure}	
	\resizebox{\hsize}{!}{\includegraphics[scale=1]{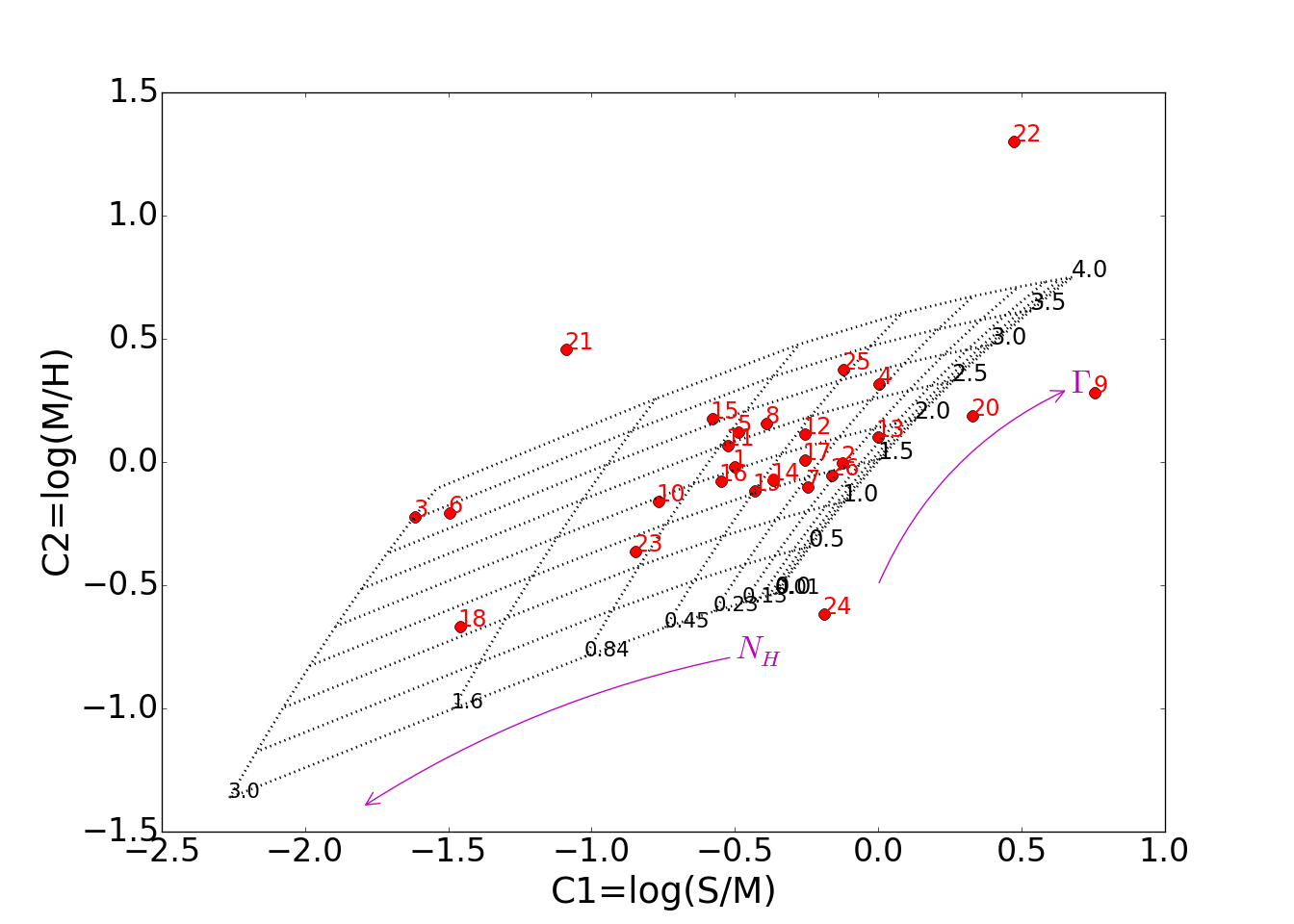}}
	\resizebox{\hsize}{!}{\includegraphics[scale=1]{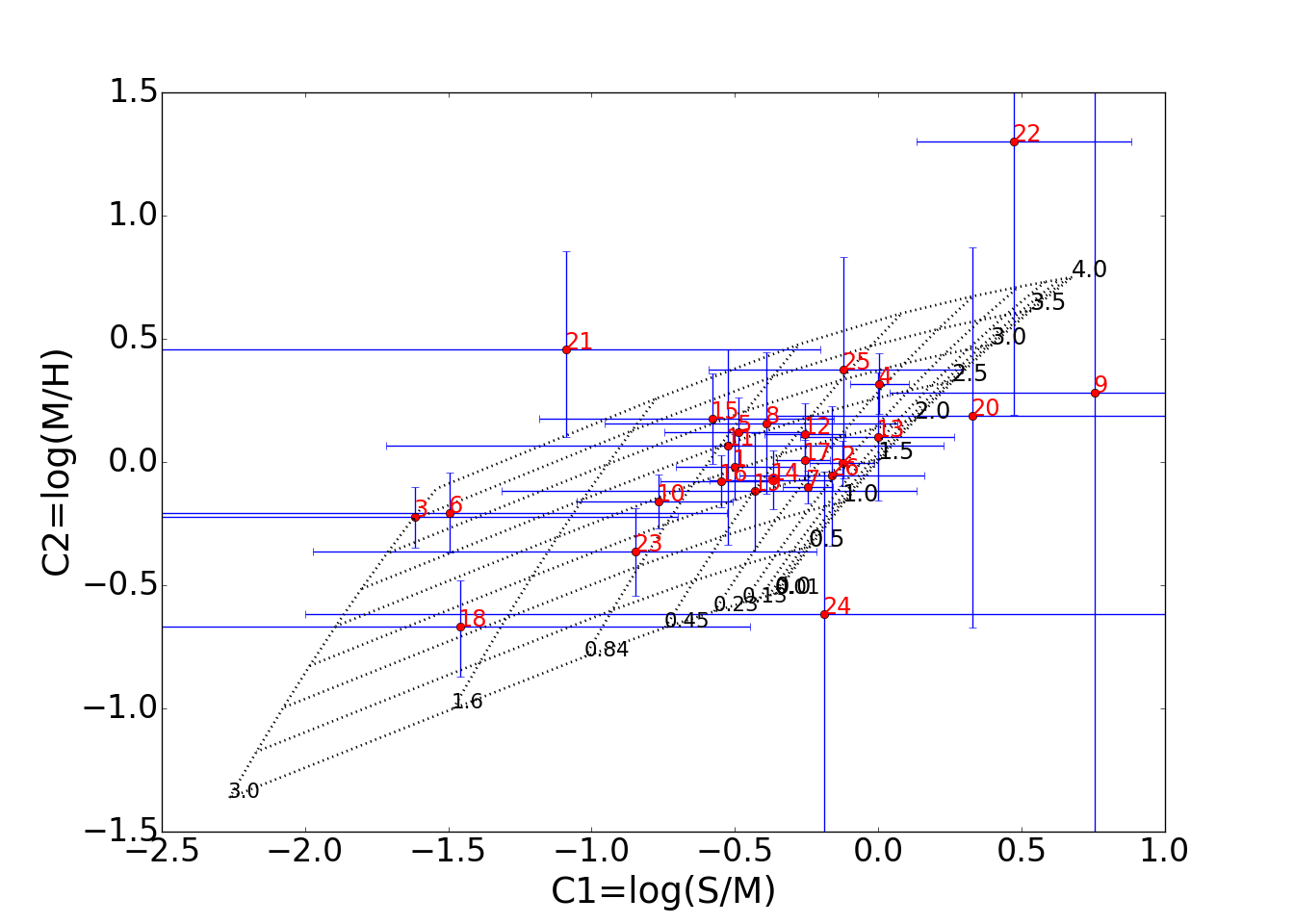}}
		\caption{(Top panel) $C_2\equiv\log_{10} ({M}/{H})$ versus $C_1\equiv\log_{10} ({S}/{M})$ X-ray colour. Overlaid is the grid of different hydrogen column densities (in units of $\mathrm{10^{22}\ cm^{-2}}$) and photon indices assuming a simple absorbed power-law model . With the red filled circles we plot our sources and with the magenta arrows we show the direction of increasing hydrogen column density and photon index. (Bottom panel) Same as top panel but showing the corresponding errors for each source with blue bars.}
	\label{fig.grid}
\end{figure}

In order to estimate the spectral parameters of the X-ray sources we created grids on plots involving two different X-ray colours and placed our sources on them. The grids were calculated by simulating absorbed power-law spectra.  Fig. \ref{fig.grid} shows our sources on the grid of the $\mathrm{C_2-C_1}$ plot. We see that while the majority of the sources fall on the grid and particularly the region corresponding to $\mathrm{\Gamma\sim 1.5-2.5}$ and $\mathrm{N_H<0.85\times 10^{22} cm^{-2}}$, there are 5 sources (9, 20, 21, 22, and 24) that fall outside the grid indicating sources with either too low absorption or too steep photon index. For three of them (9, 22, and 24) we have X-ray spectra; two are well fitted with a soft thermal-plasma model (Src 22) or a power-law component combined with a thermal plasma (Src 9), which would increase the intensity of their soft band and shift the $\mathrm{C_1}$ colour to higher values. 
Source 24 has very large errors in both $\mathrm{C_1}$ and $\mathrm{C_2}$ colours, and particularly $\mathrm{C_1}$ which is most closely related to the column density $\mathrm{N_H}$. Therefore  we decided to fix its $\mathrm{N_H}$ to the Galactic line-of-sight value. 
For the rest of the sources we estimate their spectral parameters based on their location on the grid and we present these estimates in Table \ref{tab.spectralparam}. In Table \ref{tab.spectralparam} we do not include the results of the colours based analysis for sources 7, 9, 18, and 22 which have more complex spectra.
For sources 20 and 21 that we cannot estimate their spectral parameters based on their X-ray colours we adopt in our following analysis $\mathrm{\Gamma=1.7}$ and the hydrogen column density at the galactic value $\mathrm{N_H=9.5\times 10^{19}  cm^{-2}}$ (Table \ref{tab.spectralparam}). This is consistent with the X-ray colour of Src 20 (Fig. \ref{fig.grid}), while Src 21 appears to have somewhat softer colours.

\begin{figure}	
\resizebox{\hsize}{!}{\includegraphics[scale=1]{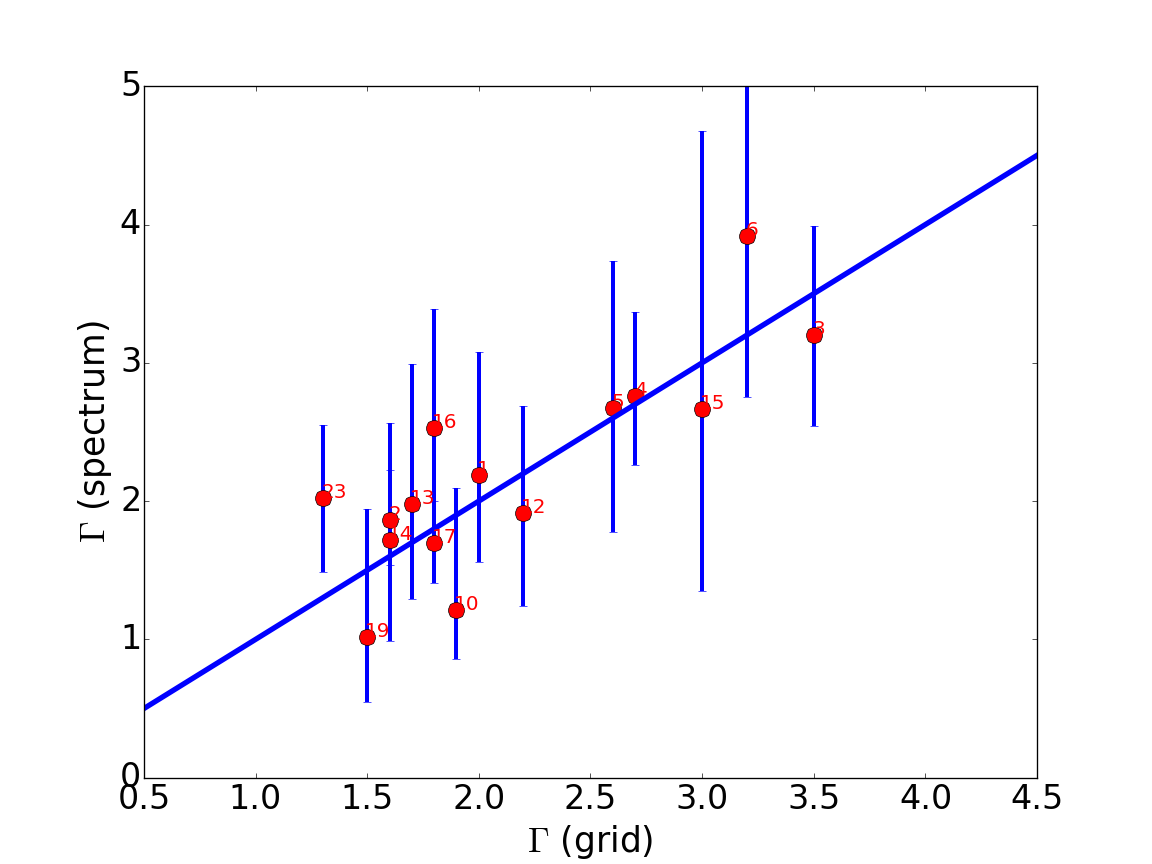}}
\caption{Photon index calculated from the spectral analysis versus photon index calculated from the X-ray colours for our sources with spectra fitted with a simple absorbed power-law model. The errors from the spectral fit are shown with blue bars. The blue solid line is the 1:1 line.}
	\label{fig.gridspectrumgamma}
\end{figure}

\begin{figure}	
\resizebox{\hsize}{!}{\includegraphics[scale=1]{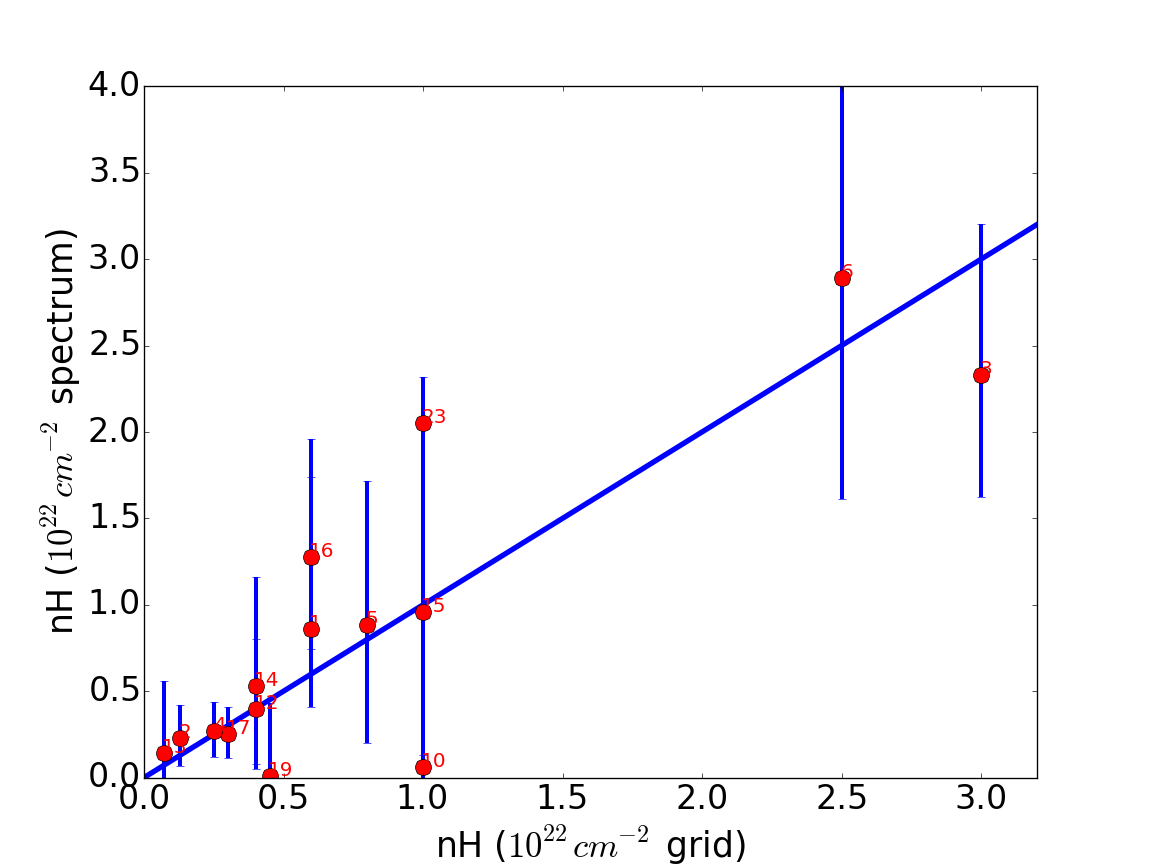}}
\caption{Hydrogen column density calculated from the spectral analysis versus hydrogen column density calculated from the X-ray colours for our sources with spectra fitted with a simple absorbed power-law model. The errors from the spectral fit are shown with blue bars. The blue solid line is the 1:1 line.}
	\label{fig.gridspectrumnh}
\end{figure}

We checked the accuracy of the spectral parameters, calculated based on their X-ray colours, by comparing them with those calculated from the spectral fit (see Table \ref{tab.spectralparam}) and we found that they agree well within their errors (Fig. \ref{fig.gridspectrumgamma} and \ref{fig.gridspectrumnh}). For the four sources (7, 9, 18, and 22) which were not fitted with a single absorbed power-law model as well as for the sources not falling on the grid (sources 20, 21, and 24) we cannot make this comparison.

\subsection{Variability}\label{variability}

We searched for variability with the \textit{glvary} tool applied on the two reprojected events files. This tool uses the Gregory-Loredo algorithm \citep{gl}.
First we ran the \textit{dither\_region} tool which creates a  normalized effective area file to be used with \textit{glvary}. The latter accounts for the fraction of the total exposure that a source falls on bad pixels or goes off the detector. For our sample the fractional area was always 1 meaning that the source was always within good detector pixels.
This ensures that any measured time variation is intrinsic and not due to the combination of the dither motion of the spacecraft and the pixel-to-pixel variation in the detector effective area. Then \textit{glvary} assigns a variability index that takes values from 0 to 10, with the value of 0 indicating a definitely not variable source and values from 6 to 10 indicating a definitely variable source. 

The results of the \textit{glvary} tool did not identify any variable source for observation OBSID 15619 with the majority of the sources being definitely not variable (variability index=0) and  some considered not variable and probably not variable (variability index=1 or 2). 

For OBSID 15077 we found that only one source showed evidence for variability: Source 19 presented a variability index of 8 indicating that it is definitely a variable source.  Inspecting its lightcurve one can see a dramatic reduction in the observed counts by about a factor of 6 in the second half of the observation. 
Also we explored its long-term variability by analysing data from the first observation of Arp\,299 (OBSID 1641), and OBSID 15619. We calculated its broad-band observed flux ($\mathrm{0.1-10.0\ keV}$) based on spectra extracted from each observation and the best-fit parameters we calculated from the merged observation. In the case of OBSID 15077 we split the observation into two segments at the time when the counts decreased. Source 19 was not detected in the first observation of Arp\,299 \citep{z03}. Nonetheless we can use BEHR \citep{park} to calculate the 90\% confidence interval upper limit on the broad-band intensity of the source even though it is not formally detected in any band.

\begin{figure}	
		\resizebox{\hsize}{!}{\includegraphics[scale=1]{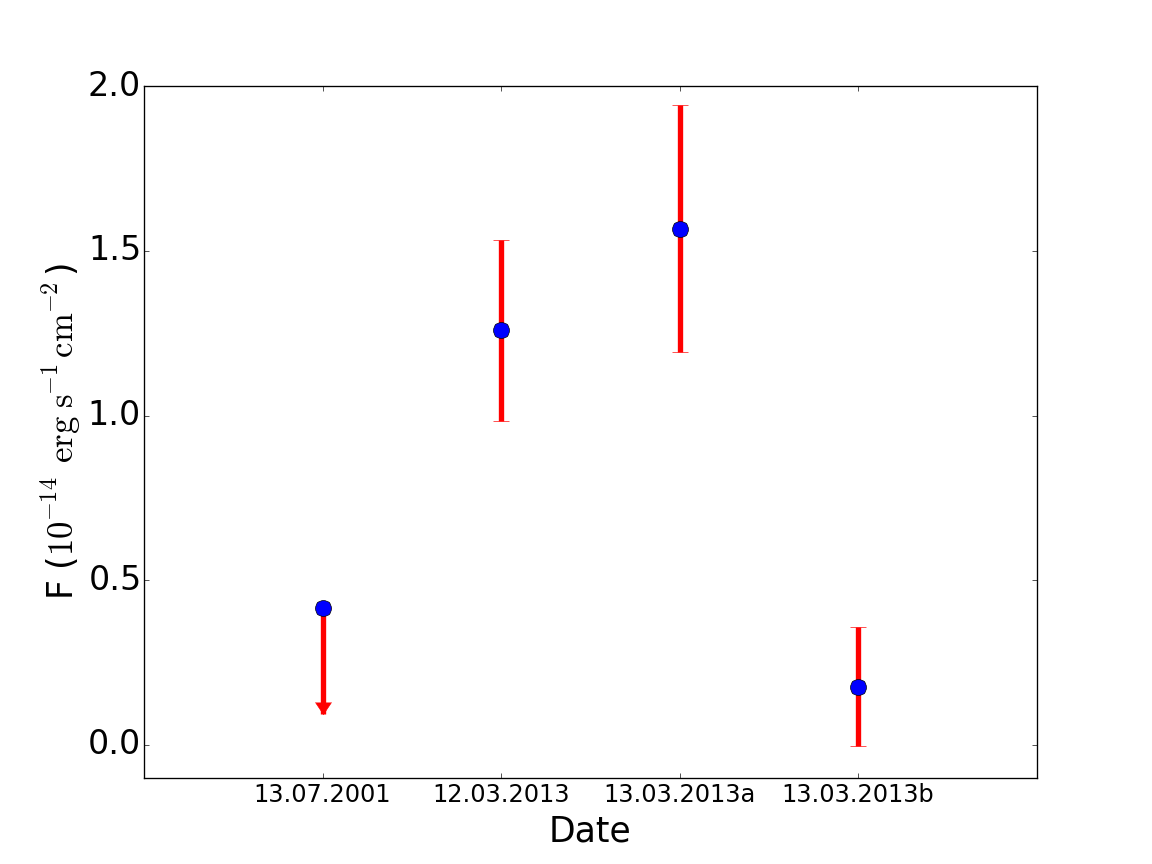}}
		\caption{Observed flux in the broad band ($\mathrm{0.1-10.0\ keV}$) versus date of observation for source 19. The corresponding errors are shown with red bars.}
	\label{fig.variable}
\end{figure}

In Fig. \ref{fig.variable} we present the broad band flux ($\mathrm{0.1-10.0\ keV}$) of source 19 versus the date of observation. We can see that the observations 15619 and 15077a do not show major differences as their fluxes agree within the errors but there is a difference with both the first observation performed in 2001 and the second half of OBSID 15077 by at least a factor of 4.

Furthermore we examined the long-term variability of all sources that were also detected at \citet{z03}. We found 6 variable sources that differed in luminosity by a factor of 3 or more; namely the sources 6, 7, 10, 14, 17, and 18.

\subsection{Extended sources}\label{extendedsources}

The relatively large distance of Arp\,299 make it subject to possible confusion effects. Therefore, we used the \textit{srcextent} tool\footnote{\url{http://cxc.harvard.edu/ciao/threads/srcextent/}}
to determine if our sample contains extended sources. The \textit{srcextent} tool calculates the sizes and associated errors of a list of sources using the Mexican Hat Optimization (MHO) algorithm \citep{houck}. A PSF is used as input to \textit{srcextent} for more reliable results. For this purpose we used ChaRT \citep{chart} to simulate the PSF and converted the output set of rays  into a pseudo-events file by projecting the rays onto the detector-plane using MARX v.5.01.1 \citep{marx}.
This analysis resulted in nine extended sources in the broad band ($\mathrm{0.1-10\ keV}$) at the 90\% confidence level. Namely the sources 3, 4, 5, 7, 20, 21, 22, 24, and 28. Sources 3, 4, 5, 18, 22, and 24 are extended in the ($\mathrm{2.0-8.0\ keV}$) band as well, following the same analysis.

\subsection{Fluxes and Luminosities of the discrete sources.}\label{fluxesandluminositiesfthediscretesources}

The calculation of the flux for the 20 sources that had spectral fitting parameters (see Table \ref{tab.spectralparam}) was done with the XSPEC package. The errors were calculated by propagating the count errors and taking into account the count-rate to flux conversion given the specific model for each source. In order to see how much is the effect of the spectral uncertainties in the measured flux, we used the \textit{flux} command  with the \textit{error} option which accounts for spectral uncertainties in the flux calculation. The errors are estimated by drawing parameter values from their best-fit distribution and calculating the flux for each set of draws. Then the resulting fluxes are ordered and the central 90\% percent is selected to give the error range. This method increases the errors on average by a factor of 2.

The fluxes for the 6 remaining sources were calculated also using the \textit{flux} command but this time the spectral parameters were fixed to the values determined based on their X-ray colours (Table \ref{tab.spectralparam}). Flux uncertainties are then simply calculated by propagating the count-rate errors reported in Table \ref{tab.properties}.  The resulting observed and absorption corrected luminosities in the broad ($\mathrm{0.1-10.0\ keV}$), soft ($\mathrm{0.1-2.0\ keV}$), and hard ($\mathrm{2.0-10.0\ keV}$) bands  are presented in Table \ref{tab.luminosities}.
 
We find that 22 sources (20 of which are off-nuclear) have luminosities above $\mathrm{10^{39}\ erg\ s^{-1}}$, (i.e the ULX limit) reaching observed luminosities of  $\mathrm{4.8\times 10^{40}\ erg\ s^{-1}}$ ($\mathrm{0.1-10\ keV}$). Our sample also includes 4 sources below the ULX limit, with the faintest source at  $\mathrm{4.0\times 10^{38}\ erg\ s^{-1}}$ ($\mathrm{0.1-10\ keV}$).

\begin{table*}
	\centering                             
	\begin{minipage}{140mm}
		\caption{Observed and absorption-corrected fluxes and luminosities.}
		\hskip-2.0cm\begin{tabular}{@{}lllllll@{}}
			\hline   
			Src\footnote{Column 1: The source identification number, ULXs are denoted with \textasteriskcentered; Columns 2, 3, and 4: Fluxes in the broad, soft, and hard bands respectively; Columns 5, 6, and 7: Luminosities in the broad, soft, and hard bands respectively.}  & $\mathrm{fx^{obs}(f_x^{corr})}$ & $\mathrm{fx^{obs}(f_x^{corr})}$  & $\mathrm{fx^{obs}(f_x^{corr})}$  &  $\mathrm{L_x^{obs}(L_x^{corr})}$ & $\mathrm{L_x^{obs}(L_x^{corr})}$   & $\mathrm{L_x^{obs}(L_x^{corr})}$  \\
			ID & ($\mathrm{0.1-10.0\ keV}$) &($\mathrm{0.1-2.0\ keV}$) & ($\mathrm{2.0-10.0\ keV}$)& ($\mathrm{0.1-10.0\ keV}$) &($\mathrm{0.1-2.0\ keV}$) & ($\mathrm{2.0-10.0\ keV}$)\\
			& $\mathrm{10^{-14}\ erg\ s^{-1}\ cm^{-2}}$  &  $\mathrm{10^{-14}\ erg\ s^{-1}\ cm^{-2}}$&  $\mathrm{10^{-14}\ erg\ s^{-1}\ cm^{-2}}$& $\mathrm{10^{39}\ erg\ s^{-1}}$     & $\mathrm{10^{39}\ erg\ s^{-1}}$  & $\mathrm{10^{39}\ erg\ s^{-1}}$    \\
			(1) & (2) & (3) & (4) & (5) & (6) & (7)\\
			\hline
			1\textasteriskcentered  &      1.6 $\pm 0.1$ (5.6)      &   0.3$\pm 0.1$ (4.2)     &  1.3 $\pm  0.2$ (1.4)   &    3.7 $\pm  0.3$ (13.0)    &    0.7$\pm 0.1$ (9.7)     &    3.0 $\pm  0.4$ (3.3)   \\
			 2\textasteriskcentered  &      4.7 $\pm 0.3$ (8.12)     &   1.4$\pm 0.1$ (4.7)     &  3.4 $\pm  0.3$ (3.4)   &    10.9$\pm  0.6$ (18.8)    &    3.2$\pm 0.2$ (10.8)    &    7.8 $\pm  0.7$ (8.0)   \\
			 3                        &      4.8 $\pm 0.4$ (240.9)    &   0.8$\pm 0.1$ (235.3)   &  4.0 $\pm  0.4$ (5.6)   &    11.1$\pm  0.9$ (558.0)   &    1.9$\pm 0.3$ (544.9)   &    9.2 $\pm  0.9$ (13.0)  \\
			 4                        &      4.4 $\pm 0.3$ (27.8)     &   2.4$\pm 0.2$ (25.7)    &  2.0 $\pm  0.3$ (2.0)   &    10.2$\pm  0.7$ (64.5)    &    5.6$\pm 0.4$ (59.7)    &    4.6 $\pm  0.7$ (4.8)   \\
			 5                        &      2.0 $\pm 0.2$ (17.6)     &   0.6$\pm 0.1$ (16.0)    &  1.45$\pm  0.2$ (1.6)   &    4.7 $\pm  0.4$ (40.9)    &    1.3$\pm 0.2$ (37.1)    &    3.4 $\pm  0.5$ (3.8)   \\
			 6\textasteriskcentered  &      1.7 $\pm 0.2$ (747.7)    &   0.3$\pm 0.1$ (745.4)   &  1.4 $\pm  0.2$ (2.2)   &    4.0 $\pm  0.5$ (1732.0)  &    0.8$\pm 0.2$ (1726.0)  &    3.2 $\pm  0.4$ (5.3)   \\
			 7                        &      27.5$\pm 0.9$ (66.4)     &   2.2$\pm 0.1$ (40.0)    &  25.3$\pm  1.2$ (26.4)  &    63.8$\pm  2.1$ (153.8)   &    5.0$\pm 0.3$ (92.7)    &    58.8$\pm  2.7$ (61.1)  \\
			 8                        &      0.3 $\pm 0.1$ (1.02)     &   0.1$\pm 0.1$ (0.7)     &  0.2 $\pm  0.1$ (0.2)   &    0.7 $\pm  0.2$ (2.3)     &    0.2$\pm 0.1$ (1.8)     &    0.5 $\pm  0.2$ (0.5)   \\
			 9\textasteriskcentered  &      1.4 $\pm 0.1$ (10.9)     &   1.0$\pm 0.1$ (10.5)    &  0.4 $\pm  0.1$ (0.4)   &    3.3 $\pm  0.1$ (25.3)    &    2.3$\pm 0.1$ (24.3)    &    0.9 $\pm  0.1$ (1.0)   \\
			 10\textasteriskcentered  &      4.2 $\pm 0.3$ (4.6)      &   0.8$\pm 0.1$ (1.2)     &  3.4 $\pm  0.3$ (3.4)   &    9.8 $\pm  0.7$ (10.7)    &    1.9$\pm 0.2$ (2.8)     &    7.9 $\pm  0.8$ (7.9)   \\
			 11                       &      0.2 $\pm 0.1$ (0.7)      &   0.1$\pm 0.1$ (0.5)     &  0.2 $\pm  0.1$ (0.2)   &    0.6 $\pm  0.1$ (1.7)     &    0.1$\pm 0.1$ (1.3)     &    0.4 $\pm  0.2$ (0.4)   \\
			 12\textasteriskcentered   &      2.2 $\pm 0.2$ (4.46)     &   0.5$\pm 0.1$ (2.7)     &  1.7 $\pm  0.2$ (1.8)   &    5.2 $\pm  0.4$ (10.3)    &    1.2$\pm 0.1$ (6.2)     &    3.9 $\pm  0.6$ (4.1)  \\
			 13\textasteriskcentered   &      0.7 $\pm 0.1$ (1.24)     &   0.3$\pm 0.1$ (0.8)     &  0.4 $\pm  0.1$ (0.4)   &    1.6 $\pm  0.2$ (2.9)     &    0.6$\pm 0.1$ (1.8)     &    1.0 $\pm  0.3$ (1.0)   \\
			 14\textasteriskcentered  &      2.7 $\pm 0.2$ (4.8)      &   0.5$\pm 0.1$ (2.4)     &  2.3 $\pm  0.3$ (2.4)   &    6.4 $\pm  0.5$ (11.0)    &    1.1$\pm 0.1$ (5.5)     &    5.3 $\pm  0.6$ (5.5)   \\
			 15\textasteriskcentered  &      1.4 $\pm 0.2$ (12.6)     &   0.4$\pm 0.1$ (11.4)    &  1.0 $\pm  0.2$ (1.2)   &    3.3 $\pm  0.4$ (29.3)    &    0.9$\pm 0.1$ (26.5)    &    2.4 $\pm  0.5$ (2.7)   \\
			 16\textasteriskcentered  &      2.4 $\pm 0.2$ (18.1)     &   0.5$\pm 0.1$ (15.8)    &  2.0 $\pm  0.2$ (2.3)   &    5.7 $\pm  0.4$ (42.0)    &    1.1$\pm 0.1$ (36.7)    &    4.6 $\pm  0.5$ (5.4)   \\
			 17\textasteriskcentered  &      7.3 $\pm 0.4$ (11.1)     &   1.7$\pm 0.1$ (5.4)     &  5.6 $\pm  0.5$ (5.7)   &    16.9$\pm  0.8$ (25.8)    &    4.0$\pm 0.3$ (12.5)    &    12.9$\pm  1.1$ (13.2)  \\
			 18                       &      13. $\pm 1.2$ (18.0)     &   0.3$\pm 0.1$ (3.2)     &  13.2$\pm  0.9$ (14.7)  &    31.4$\pm  2.7$ (41.7)    &    0.8$\pm 0.2$ (7.4)     &    30.6$\pm  2.3$ (34.2)  \\
			 19\textasteriskcentered  &      1.0 $\pm 0.2$ (1.1)      &   0.2$\pm 0.1$ (0.2)     &  0.9 $\pm  0.2$ (0.9)   &    2.5 $\pm  0.4$ (2.6)     &    0.5$\pm 0.1$ (0.5)     &    2.1 $\pm  0.5$ (2.1)   \\
			 20                       &      0.5 $\pm 0.1$ (0.5)      &   0.2$\pm 0.1$ (0.2)     &  0.2 $\pm  0.1$ (0.2)   &    1.2 $\pm  0.3$ (1.2)     &    0.5$\pm 0.2$ (0.6)     &    0.4 $\pm  0.2$ (0.4)   \\
			 21                       &      0.5 $\pm 0.1$ (0.5)      &   0.2$\pm 0.1$ (0.2)     &  0.3 $\pm  0.1$ (0.3)   &    1.2 $\pm  0.3$ (1.2)     &    0.5$\pm 0.1$ (0.5)     &    0.7 $\pm  0.3$ (0.7)   \\
			 22                       &      1.8 $\pm 0.1$ (13.7)     &   1.6$\pm 0.1$ (13.4)    &  0.2 $\pm 0.1$ (0.2)   &    4.2 $\pm  0.2$ (31.8)    &    3.7$\pm 0.2$ (31.2)    &    0.5 $\pm 0.1$ (0.5)   \\
			 23\textasteriskcentered   &      2.4 $\pm 0.3$ (8.12)     &   0.2$\pm 0.1$ (5.4)     &  2.3 $\pm  0.3$ (2.7)   &    5.6 $\pm  0.6$ (18.8)    &    0.4$\pm 0.1$ (12.5)    &    5.2 $\pm  0.7$ (6.4)   \\
			 24                       &      0.7 $\pm 0.2$ (10.5)     &   0.1$\pm 0.1$ (9.6)     &  0.7 $\pm  0.1$ (0.9)   &    1.6 $\pm  0.4$ (24.4)    &    0.1$\pm 0.1$ (22.2)    &    1.5 $\pm  0.3$ (2.2)   \\
			 25                       &      0.1 $\pm 0.1$ (0.9)      &   0.1$\pm 0.1$ (0.9)     &  0.1 $\pm  0.1$ (0.1)   &    0.4 $\pm  0.1$ (2.2)     &    0.2$\pm 0.1$ (2.0)     &    0.2 $\pm  0.1$ (0.2)   \\
			 26\textasteriskcentered   &      0.5 $\pm 0.1$ (0.6)     &   0.1$\pm 0.1$ (0.2)    &  0.4 $\pm  0.1$ (0.4)   &    1.3 $\pm  0.3$ (1.4)     &    0.3$\pm 0.1$ (0.4)     &    1.0 $\pm  0.3$ (1.0)   \\
			 \hline
		\end{tabular} 	
		\label{tab.luminosities}
			\end{minipage}
		\end{table*}

\subsection{Integrated X-ray emission of the galaxy}\label{integratedxrayemissionfthegalaxy}

\begin{figure}
	\resizebox{\hsize}{!}{\includegraphics[scale=1,angle=270]{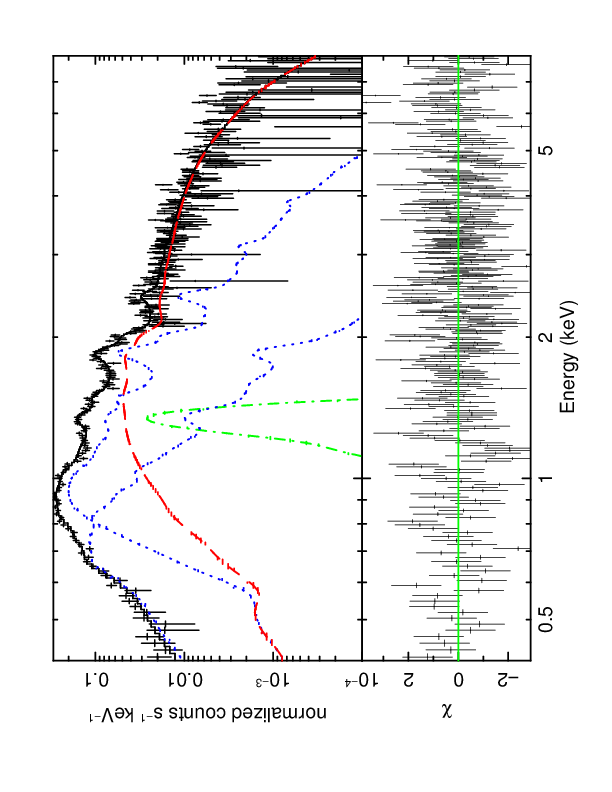}}
	\caption{(Top panel) The integrated X-ray spectrum of Arp\,299, along with the best-fit folded model consisting of: An absorbed power-law component (dashed red line), two apec components (dotted blue lines), and a Gaussian line (dashed-dotted green line). (Bottom panel) The fit residuals in terms of sigmas with error bars of size 1$\mathrm{\sigma}$.}		
	\label{fig.spectotallx}
\end{figure}

We extracted the integrated spectrum of the galaxy encompassed within its D25 area \citep[RC3;][]{rc31,rc32}. The background spectrum was determined from a source-free area outside the D25 area using the \textit{specextract} tool. Based on previous observations of this galaxy \citep{z03} which indicate the presence of X-ray binaries together with a hot gaseous component, we obtained the best-fit ($\mathrm{\chi_{\nu}^2/dof=331.88/294}$) with a model consisting of an absorbed thermal plasma component and a power-law plus a second thermal component, both seen through an additional absorber. The first absorption component was fixed to the Galactic line-of-sight value. We also added a Gaussian line to account for a line-like feature at 1.3 keV which significantly improves the fit without affecting the values of any other model parameters (Fig \ref{fig.spectotallx}). Therefore the full model in XSPEC was: \texttt{phabs(gaussian+apec+phabs(apec+ powerlaw))}.

The best-fit parameters are $\Gamma=1.67_{-0.12}^{+0.11}$, $\mathrm{N_{H_1}=9.5\times 10^{19}\ cm^{-2}}$, $\mathrm{N_{H_2}=(0.55_{-0.07}^{+0.06})\times 10^{22}\ cm^{-2}}$, $\mathrm{kT_1=0.32\pm 0.01\ keV}$ and $\mathrm{kT_2=0.83_{-0.03}^{+0.04}\ keV}$ (Table \ref{tab.specparam_ulxs}).
These results are consistent with studies of the integrated X-ray spectra of other star-forming galaxies \citep[e.g.][]{z98,cappi,moran,pereira}.

The results of the absorbed, corrected for the Galactic absorption, and unabsorbed luminosities in the broad ($\mathrm{0.1-10.0\ keV}$), soft ($\mathrm{0.1-2.0\ keV}$), and hard ($\mathrm{2.0-10.0\ keV}$) bands are shown in Table \ref{tab.binulxslum}. 

\subsubsection{Luminosity of X-ray binaries and ULXs}\label{luminosityofxraybinariesandulxs}

X-ray binaries are the dominant component of X-ray emission of galaxies above $\mathrm{2\ keV}$ \citep[e.g.][]{lira, fabbianorev}. We measure the integrated emission of both resolved and unresolved XRBs in Arp\,299 by extracting the integrated spectrum of the galaxy, excluding Src 7 which corresponds to the AGN in NGC 3690 (\S \ref{nucleusofngc3690}). We fitted this spectrum, with a model consisting of an absorbed thermal plasma component and a power-law plus a second thermal component, both seen through an additional absorber (full model in XSPEC: \texttt{phabs(apec+phabs(apec+ powerlaw))}; $\mathrm{\chi^2=302.11/282}$). The power-law component is then attributed to the total population of X-ray binaries which may be embedded in a star-forming region, hence requiring an additional absorption. In this way we account not only for the discrete X-ray sources but also for the unresolved X-ray binaries. The best-fit parameters for this fit are presented in \ref{tab.specparam_ulxs} (second row) and the corresponding absorbed and unabsorbed luminosities in Table \ref{tab.binulxslum}.

As mentioned previously ULXs are off-nuclear point sources that have luminosities $\mathrm{L_{(0.1-10.0\ keV)}>10^{39}\ erg\ s^{-1}}$. In our sample we found 20 off-nuclear sources that fulfil the criterion and belong to the ULX class. In these we do not include Src 22 which had a soft thermal plasma spectrum that is inconsistent with an X-ray binary.  If we account only for the point sources in the broad-band our sample reduces to 14 ULXs (denoted with an asterisk in Table \ref{tab.luminosities}).

Given the distance of Arp\,299 we expect that this observation suffers from source confusion which affects the number of observed sources. In order to avoid this complication our analysis focuses on the total luminosity of ULXs which is not affected by source confusion. When we discuss the source numbers, we give results for the total ULX sample (20 sources), as well as for the more conservative choice of point-like objects (14 sources).

We calculated the total flux of the 20 and 14 ULXs by extracting a spectrum from a region consisting of all their apertures and using a background spectrum from a source-free area outside the D25 area. A model consisting of an absorbed power-law and a thermal-plasma model with variable abundances gave an acceptable fit but with significant residuals in the $\mathrm{1.0-2.0\ keV}$ range (Model in XSPEC: \texttt{phabs(vapec+po)}). Therefore, we also used an absorbed thermal-plasma (with variable abundances) and a disk blackbody model \citep{mitsuda,makishima} which has been successfully used to model the spectra of ULXs \citep{gladstone,rana} (Model in XSPEC: \texttt{phabs(vapec+diskbb)}). The best-fit parameters for both models are shown in Table \ref{tab.specparam_ulxs} and they are in good agreement with fits to the \textit{XMM-Newton} spectra of nearby ULXs \citet{gladstone}. 

\begin{table*}
	\centering
	\begin{minipage}{140mm}
			\caption{Spectral fitting parameters of integrated spectrum of galaxy, binaries, ULXs, and diffuse emission}
		\hskip-2.6cm
		\begin{tabular}{@{\extracolsep{1pt}}llllllllll@{}}
			\hline 			
			 & \multicolumn{3}{l}{Power-law} & \multicolumn{3}{l}{Thermal plasma } &\multicolumn{2}{l}{Disk blackbody} & \\[5pt]
			\cline{2-4}
			\cline{5-7}
			\cline{8-9}
			&&&&&&&&&\\[-5pt]
			Region & $\mathrm{N_H}$ & $\Gamma$ & Norm\footnote{Normalization of power-law component in units of $\mathrm{10^{-5}\ photons\ keV^{-1}\ cm^{-2}\ s^{-1}}$ at 1 keV.} & kT & Z& Norm\footnote{Normalization of thermal plasma component in units of $\mathrm{10^{-5}\ photons\ keV^{-1}\ cm^{-2}\ s^{-1}}$ at 1 keV.}  & Tin & Norm\footnote{Normalization of disk black body component in units of $\mathrm{10^{-3}\ photons\ keV^{-1}\ cm^{-2}\ s^{-1}}$ at 1 keV.}& \\
			& $10^{22}\ cm^{-2}$ & & & keV  &  $(\times\mathrm{Z_{\odot}})$ & & keV       &     &    $\mathrm{\chi^2/dof}$ \\ 
			\hline
			\hline
			Total galaxy\footnote{Also includes a component to account for the Galactic line-of-sight absorption ($\mathrm{N_H=9.5\times 10^{19}  cm^{-2}}$).} & $0.55_{-0.07}^{+0.06}$ & $1.67_{-0.12}^{+0.11}$& $26.33_{-3.94}^{+4.03}$ & $0.32\pm 0.01$ &-&$13.4\pm 1.0$&-&-& 331.88/294\\[5pt]
			 &  & &&$0.83_{-0.03}^{+0.04}$&&$70.1\pm 14.0$&&&\\[10pt]	
			 Total galaxy (no AGN)\footnote{Also includes a component to account for the Galactic line-of-sight absorption ($\mathrm{N_H=9.5\times 10^{19}  cm^{-2}}$ ).} & $0.55\pm 0.06$ & $1.80_{-0.13}^{+0.12}$ & $27.17_{-4.45}^{+4.38}$ & $0.32\pm 0.01$ &- &$13.4\pm 1.0$ &- & -&302.11/282 \\ [5pt]
			 &&&&  $0.84\pm 0.03$ &&$67.3\pm 14.3$&&&\\[10pt]	
			 All ULXs (20) &  $0.27_{-0.07}^{+0.05}$ & - &  -& $0.62_{-0.11}^{+0.14}$ & $0.18_{-0.06}^{+0.08}$ (Fe) & $11.0_{-3.5}^{+3.8}$& $1.42_{-0.09}^{+0.11}$ & $6.6_{-1.6}^{+2.0}$ & 160.75/164\\[5pt]
			&$0.50_{-0.08}^{+0.12}$&$2.04_{-0.10}^{+0.12}$&$18.1_{-2.2}^{+3.0}$ &$0.31_{-0.06}^{+0.08}$&$0.71_{-0.30}^{+0.53}$ (Fe)&$19.2_{-10.6}^{+28.1}$&-&-&176.00/164\\[10pt]		
			Point-like ULXs (14)  & $0.23_{-0.06}^{+0.12}$ &- &- & $0.54_{-0.15}^{+0.14}$ & $0.19_{-0.09}^{+0.21}$ (Fe) & $4.1_{-2.2}^{+4.0}$ & $1.51_{-0.14}^{+0.13}$ & $3.4_{-0.9}^{+1.67}$ &  121.60/131\\[5pt]
			&$0.55\pm 0.14$&$1.95_{-0.12}^{+0.11}$&$10.9_{-1.6}^{+1.7}$&$0.24_{-0.04}^{+0.06}$&1.00 (Fe)&$20.0_{-15.2}^{+47.2}$&-&-&130.08/132\\[10pt]		
			Diffuse emission & $7.45\pm 0.02$& $1.52_{-0.24}^{+0.21}$& $7.0_{-1.8}^{+2.0}$ & $0.72\pm 0.03$ &$1.24_{-0.29}^{+0.34}$ (Ne)& $49.0_{-7.4}^{+7.8}$&-&-& 201.44/210\\[5pt]	
			&&&&&$1.14_{-0.20}^{+0.25}$ (Mg) &&&&\\[5pt]
			&&&&& $0.26_{-0.03}^{+0.04}$ (Fe)&&&&\\		
			\hline
		\end{tabular} 	
		\label{tab.specparam_ulxs}
	\end{minipage}
\end{table*}

In order to measure the integrated luminosity of the ULXs, we used the second model (\texttt{phabs(vapec+diskbb)}) because it gave a marginally better fit and accounted only for the \textit{diskbb} component. The thermal component is interpreted as the contribution of the diffuse emission of the galaxy which could not be subtracted completely by the background spectrum. This is supported by the fact that the abundance of Fe is in agreement with that of the diffuse emission (see \S \ref{luminosityofthediffuseemission}). The integrated absorbed and unabsorbed luminosities of the combined 20 and the combined 14 ULXs are reported in Table \ref{tab.binulxslum}.

\subsubsection{Luminosity of the diffuse emission}\label{luminosityofthediffuseemission}

\begin{figure}	
	\resizebox{\hsize}{!}{\includegraphics[scale=1]{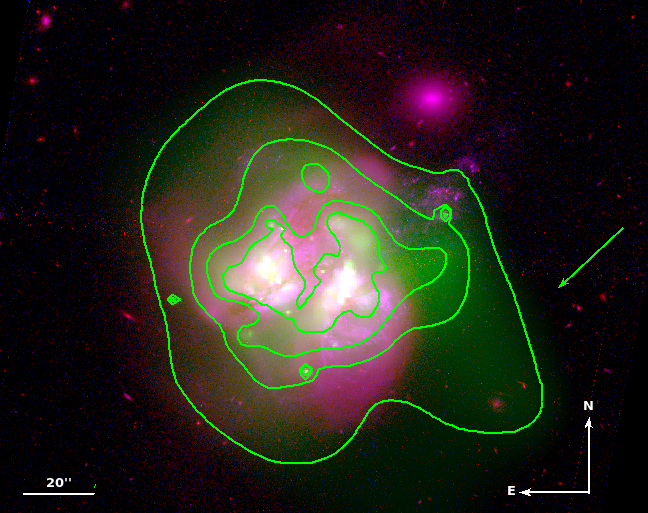}}
	\caption{Composite colour image of Arp\,299 where the soft smoothed image at $\mathrm{0.1-1.2\ keV}$ is the green colour and red and blue colours are the F814W ($\mathrm{814\ nm}$) and F435W ($\mathrm{435\ nm}$) images from \textit{HST}. The contours outline the distribution of the diffuse X-ray emission and the green arrow indicates the location of the plume in the west of the galaxy.}
	\label{fig.superwind}
\end{figure}

The diffuse emission of star-forming galaxies shows evidence for a hot thermal gaseous component often associated with a galactic scale superwind  \citep[e.g.][]{strickland}. Arp\,299 is not an exception as seen from fits to its integrated X-ray spectrum (\S \ref{integratedxrayemissionfthegalaxy}). Furthermore, optical observations of Arp\,299 show evidence for a galactic-scale wind \citep{b3}. A plume in the west of NGC 3690 (Fig. \ref{fig.superwind}) extending beyond the optical outline of the galaxy would be associated with such a superwind.

In order to measure the luminosity of the diffuse soft ($\mathrm{0.1-2.0\ keV}$) and hard ($\mathrm{2.0-10.0\ keV}$) X-ray emission of Arp\,299, we created ``swiss cheese'' images in each of the two bands after removing the regions corresponding to the 25 detected X-ray sources.
To obtain a picture of its spectral parameters we extracted a spectrum from the D25 region of the galaxy (also excluding the discrete X-ray sources) and a background spectrum from a source-free area outside the D25 area using the \textit{specextract} tool. The spectrum was fitted with an absorbed power-law model (to account for unresolved X-ray binaries) plus a thermal plasma model with variable abundances. We also added a second photoelectric absorption component fixed to the line-of-sight Galactic absorption (model in XSPEC: \texttt{phabs(phabs(po+vapec))}). The abundances of Ne, Mg, and, Fe were left free to vary. For the other elements we adopted solar abundances. We tried to thaw the oxygen and silicon elements but since there was no improvement in the fit, and their resulting values were close to the solar, we adopted a solar value for these elements as well. The best-fit parameters for this model ($\mathrm{\chi^2_{\nu}/dof=201.44/210}$) and their corresponding errors at the 90\% confidence level are shown in Table \ref{tab.specparam_ulxs}.

Since this spectrum does not account for the diffuse emission that lies within the excised source regions we cannot directly compute the flux and therefore the luminosity of the diffuse emission of Arp\,299. For this reason we created an image of the diffuse emission by interpolating the pixel values in the source regions with values from annular regions surrounding them, using the \textit{dmfilth} tool. These annular regions were the same as those used as background regions in the spectral analysis. Then we measured the total diffuse emission intensity within the D25 region of the galaxy. Using the spectral model and making the implicit assumption that the spectrum in the source regions is on average the same as in the rest of the galaxy, we calculated the flux and the corresponding luminosity of the diffuse emission in the soft and hard bands by rescaling the model-predicted fluxes by the ratio of the counts in the interpolated image and the swiss-cheese image in each band.
 
The absorbed as well as the corrected luminosities for both absorption components are reported in Table \ref{tab.binulxslum}. Correcting for the Galactic line-of-sight absorption essentially makes no difference in the absorption-corrected luminosity.

\begin{table*}
	\centering
	\begin{minipage}{140mm}
		\caption{Luminosities of the integrated spectrum of the total galaxy, X-ray binaries, ULXs, and the diffuse emission}
		\hskip-3.2cm
		\begin{tabular}{@{\extracolsep{-8pt}}ccccc@{}}
			\hline    			
			N   &  $\mathrm{L_x^{obs}(L_x^{corr})}$ & $\mathrm{L_x^{obs}(L_x^{corr})}$   & $\mathrm{L_x^{obs}(L_x^{corr})}$  & $\mathrm{L_x^{obs}(L_x^{corr})}$ \\
			& ($\mathrm{0.1-10.0\ keV}$) &($\mathrm{0.1-2.0\ keV}$) & ($\mathrm{2.0-10.0\ keV}$) & ($\mathrm{0.5-8.0\ keV}$) \\
			&  $\mathrm{10^{40}\ erg\ s^{-1}}$     & $\mathrm{10^{40}\ erg\ s^{-1}}$  & $\mathrm{10^{40}\ erg\ s^{-1}}$ &    $\mathrm{10^{40}\ erg\ s^{-1}}$ \\
			\hline  	
			\hline 
			Total galaxy\footnote{The displayed luminosities for Apr\,299 are given in the form $\mathrm{L_x^{obs}(L_{xGal}^{corr}|L_x^{corr})}$ where the $\mathrm{L_x^{obs}}$ is the observed luminosity, $\mathrm{L_{xGal}^{corr}}$ is the luminosity corrected only for the Galactic line-of-sight absorption, and $\mathrm{L_x^{corr}}$ is the unabsorbed luminosity.}  & $49.0 \pm 0.4\ (51.0 \pm 0.4\ |120.0 \pm 0.8)$ & $23.0\pm 0.2 \ (23.0 \pm 0.2\ |92.0 \pm 0.8)$ & $26.0\pm 0.6 \ (27.0 \pm 0.6\ |28.0 \pm 0.6)$ & $42.1\pm 0.3\ (42.7 \pm 0.3\ |89.8 \pm 0.6)$\\
			Binaries\footnote{Includes also emission from unresolved sources.} & 25.0$\pm 0.2$ ($48.0 \pm 0.4$) & 4.5$\pm 0.4$ ($26.2 \pm 2.3$) &  21.0$\pm 0.8$ ($21.8 \pm 0.8$) & 21.7$\pm 0.2$ ($32.4 \pm 0.3$) \\ 
			ULXs20 & 10.4$\pm 0.1$ ($13.1 \pm 0.1$) & 2.6$\pm 0.4$ ($5.1 \pm 0.8$) & 7.8$\pm 0.2$ ($8.1 \pm 0.2$) &  10.0$\pm 0.1$ ($11.9 \pm 0.1$)   \\	 
			ULXs14 &  7.1$\pm 0.1$ ($8.8 \pm 0.1$)  &  1.7$\pm 0.4$ ($3.2 \pm 0.7$)   & 5.4$\pm 0.1$ ($5.5\pm 0.1$) &  6.8$\pm 0.1$ ($7.9\pm 0.1$)  \\			
			Diffuse emission & 22.1$\pm 0.2$ ($24.2\pm 0.2$) & 15.0$\pm 0.2$ ($17.0\pm 0.2$)& 7.1$\pm 0.7$ ($7.2\pm 0.7$) & 19.7 $\pm 0.2$ ($20.6\pm 0.2$)\\	
			\hline	
		\end{tabular} 	
		\label{tab.binulxslum}
	\end{minipage}
\end{table*}

\subsection{Nucleus of NGC 3690}\label{nucleusofngc3690}

The NGC 3690 nucleus has been extensively studied in the X-ray regime. The fist observational evidence of a heavily obscured AGN existing in Arp\,299 was revealed with \textit{Beppo-SAX} \citep{b9}. \textit{Chandra} and \textit{XMM-Newton}, as well as recent results from simultaneous observations with \textit{NuSTAR} and \textit{Chandra}, confirmed the existence of an AGN in NGC 3690 with the detection of a strong 6.4 keV line \citep{z03,ballo,ptak}.
With the excellent spatial resolution  and this deep exposure of \textit{Chandra} we are able to individually study the nucleus of NGC 3690 and its contribution in the X-ray output of Arp\,299. Therefore we extracted the spectrum of Src 7 (nucleus of NGC 3690, B1) from the same region and using the same background as in our photometric analysis (\S \ref{sourcedetection}).

We first tried a simple model consisting of an absorbed power-law, a Gaussian line and a thermal plasma component (model in XSPEC: \texttt{phabs(apec+po+gaussian)}). The best-fit parameters ($\mathrm{\chi^2_{\nu}/dof=60.92/43}$) and their corresponding uncertainties are: $\mathrm{N_{H}=(1.15\pm 0.18)\times 10^{22} cm^{-2}}$ for the hydrogen column density and $\Gamma=0.19_{-0.70}^{+0.60}$ for the photon index of the power-law component. The thermal plasma model has a temperature of $\mathrm{kT=0.55_{-0.10}^{+0.11}\ keV}$ and the Gaussian line energy is $\mathrm{E_l=6.33\pm 0.04\ keV}$ with $\mathrm{\sigma=0.10_{-0.10}^{+0.09}\ keV}$. 

\begin{figure}	
		\resizebox{\hsize}{!}{\includegraphics[scale=1,angle=270]{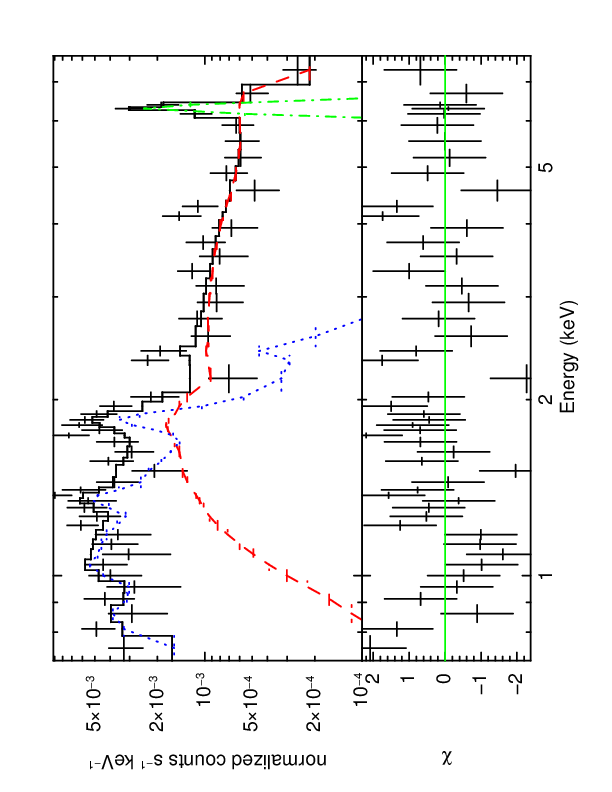}}
	\caption{(Top panel) The X-ray spectrum of the nucleus of NGC 3690 (Src 7), along with the best-fit folded model consisting of: An absorbed power-law component (dashed red line), a thermal plasma component (dotted blue line), and a Gaussian line (dashed-dotted green line). (Bottom panel) The fit residuals in terms of sigmas with error bars of size 1$\mathrm{\sigma}$.}
	\label{fig.spec_agn_test14b}
\end{figure}

We notice that $\Gamma$ is very flat indicative of a heavily obscured AGN. Therefore we try a model appropriate for heavily obscured AGN \citep[e.g.][]{b11}. This model consisted of a power-law plus a Gaussian line at $\mathrm{6.4\ keV}$ seen through a partial covering absorber (Fig. \ref{fig.spec_agn_test14b}) and a thermal-plasma component. All the components were absorbed by photoelectric absorption (model in XSPEC: \texttt{phabs(apec+pcfabs(po+gaussian))}). The adopted model corresponds to a physical picture where the emission of a deeply embedded AGN, and possibly some X-ray binaries (which we cannot isolate from the AGN), escapes and is later absorbed by another concentration of cold gas. 

We obtained a very good fit ($\mathrm{\chi^2_{\nu}/dof=52.53/41}$) where the best-fit parameters and their corresponding uncertainties (also shown in Table \ref{tab.spectralparam}) are: $\mathrm{N_{H}=(1.02_{-0.27}^{+0.22})\times 10^{22} cm^{-2}}$ for the hydrogen column density of the overall photoelectric absorption component, while for the HI column density and covering fraction the of the partial covering absorber we have $\mathrm{N_{H_{fr}}=(127.0_{-61.9}^{+51.9})\times 10^{22}\ cm^{-2}}$ and $\mathrm{f=0.95}$ (i.e. 5\% of the emission escapes the first absorber). The photon index of the power-law component is $\Gamma=1.02_{-0.75}^{+0.66}$. The still flat photon index originates probably from the fact that we are not recovering the intrinsic power-law spectrum because of the limited band of \textit{Chandra}. The thermal plasma model has a temperature of $\mathrm{kT=0.52_{-0.18}^{+0.09}\ keV}$. The Gaussian line energy is $\mathrm{E_l=6.30\pm 0.03\ keV}$ with $\mathrm{\sigma=0.018_{-0.018}^{+0.093}\ keV}$ and an equivalent width of $\mathrm{EW=0.72\pm 0.01\ keV}$. These results are consistent with the \textit{NuSTAR} spectral fits \citep{ptak} apart from the power-law photon index which in our case is much flatter than in the \textit{NuSTAR} fits. This is understandable given the much narrower energy range of \textit{Chandra}.

Table \ref{tab.agn} shows the observed and absorption corrected luminosity of the AGN. We note here that with the \textit{Chandra} data we do not recover the intrinsic AGN and therefore the luminosities we measure are only a lower limit. Nonetheless they are very useful for constraining its contribution in the observed total X-ray output of Arp\,299 in the \textit{Chandra} band. Rows 1, 2, and 3 present the absorbed luminosity, the unabsorbed luminosity corrected only for the Galactic column density, and the total unabsorbed luminosity for the broad, soft, and hard bands. Row 4 presents the AGN unabsorbed luminosity not accounting for the thermal-plasma emission in the same bands. This luminosity can give us an idea of the emission of the AGN. In all cases we cannot disentangle contamination from X-ray binaries within the AGN aperture. However the large X-ray luminosity (even if not corrected for absorption) suggests that the measured emission is dominated by the AGN.

The AGN contributes to 13.1\% of the total absorbed luminosity of Arp\,299 in the broad band ($\mathrm{0.1-10.0\ keV}$), 2.3\% in the soft band ($\mathrm{0.1-2.0\ keV}$), and 22.1\% in the hard band ($\mathrm{2.0-10.0\ keV}$). Correcting the spectrum for the photoelectric absorption (\texttt{phabs}) the contributions are 12.7\% in the broad, 10.0\% in the soft and 21.7\% in the hard band. When correcting for the second absorption (\texttt{pcfabs}) the hard X-ray luminosity of the AGN is larger than the hard X-ray luminosity of the whole galaxy. This result may seem at odds but can be explained by the fact that the contribution of the AGN in the integrated spectrum of the galaxy is only 22\%. As a result we cannot recover the AGN spectral parameters from the integrated galaxy spectrum. Furthermore, the inferred AGN luminosity is strongly model dependent on its spectral parameters, which have rather large uncertainties. Nonetheless, the AGN has only minor contribution to the observed X-ray output of the system, but it may dominate its energetics.

\begin{table*}
	\centering
	\begin{minipage}{140mm}
		\caption{AGN luminosities}
		\begin{tabular}{@{}cccc@{}}
			\hline    			
			Ltype &  L ($\mathrm{0.1-10.0\ keV}$)   &L ($\mathrm{0.1-2.0\ keV}$)  & L ($\mathrm{2.0-10.0\ keV}$)   \\
			&  $\mathrm{10^{39}\ erg\ s^{-1}}$    &   $\mathrm{10^{39}\ erg\ s^{-1}}$   &   $\mathrm{10^{39}\ erg\ s^{-1}}$     \\
			\hline
			absorbed& 63.8$\pm 2.1$ & 5.0$\pm 0.3$ & 58.8$\pm 2.7$ \\
         Galactic corrected\footnote{Corrected luminosity for the Galactic line-of-sight.} &  153.8$\pm 5.1$  &   92.7 $\pm 5.5$  &   61.1$\pm 2.8$   \\
			unabsorbed\footnote{Corrected luminosity for both photoelectric absorption components.} & 759.7$\pm 24.3$ & 211.8$\pm 12.7$ & 547.9$\pm 25.1$\\
	      AGN unabsorbed\footnote{Corrected luminosity for both photoelectric absorption and for thermal plasma components.} & 672.3 $\pm 22.1$& 125.3 $\pm 7.5$& 547.0$\pm 25.1$ \\
			\hline
		\end{tabular} 	
		\label{tab.agn}
	\end{minipage}
\end{table*}

\subsection{Nucleus of IC 694}\label{nucleusofic694}  

The nucleus of IC 694 (Arp\,299-A) has also been studied extensively and its nature is still debated. \citet{z03} with a relatively short exposure claimed that the IC 694  can be explained with a significant number of high-mass X-ray binaries (HMXBs; $5\times 10^5$ O-type stars) although the presence of a relatively weak, mildly obscured, AGN cannot be excluded. \textit{Chandra} and \textit{XMM-Newton} observations detected ionized Fe-K emission in IC 694. Part of it may come from an AGN although significant contribution may be accounted to star formation  
\citep{ballo}. Evidence of an AGN in IC 694 based on the existence of a strong flat spectrum radio source is presented in \citet{perez}. \citet{ah13} also supported the existence of both a more obscured and much less luminous AGN than in NGC 3690 by modelling its  mid-infrared spectrum. \citet{ptak} with a simultaneous observation of \textit{Chandra} and \textit{NuSTAR} showed that the lack of significant emission above $\mathrm{10\ keV}$ suggests that any AGN must be highly obscured or have a much lower luminosity than that of NGC 3690. In general the observed He-like Fe-K$\alpha$ emission line and the lack of a $\mathrm{6.4\ keV}$ Fe-K$\alpha$ line makes the presence of an AGN inconclusive. For example several nearby LIRGs/ULIRGS with no apparent AGN signatures, show strong ionized Fe-line \citep{iwasawa05,b11}, which can be accounted for by thermal emission originating from a starburst \citep[e.g.][]{strickland09}.

With the spatial resolution and high S/N of the \textit{Chandra} observation considered here we can locate the source of the Fe-K line and separate the nuclear (possible AGN) from the circumnuclear (XRBs) hard component.

\begin{figure}	
 	\resizebox{\hsize}{!}{\includegraphics[scale=1,angle=270]{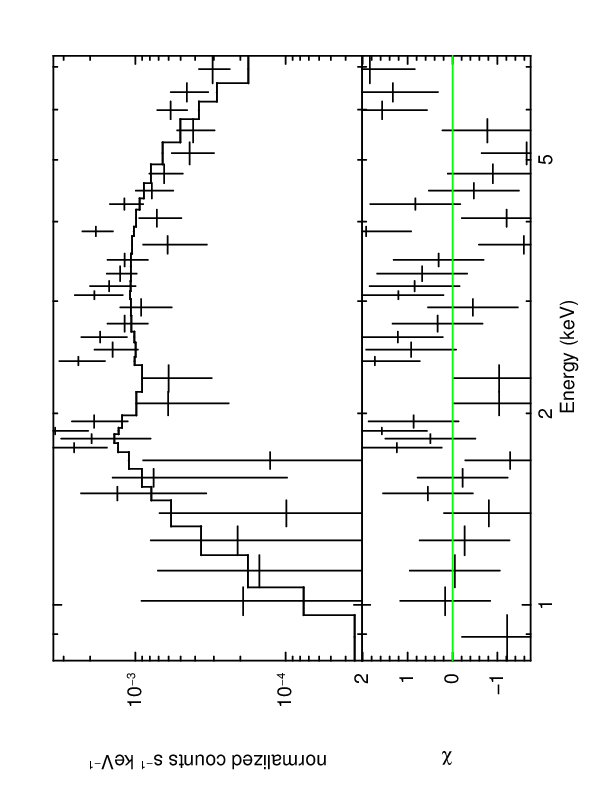}}
 	\caption{(Top panel) The X-ray spectrum of the nucleus of IC 694 (Src 18), along with the best-fit folded model consisting of an absorbed power-law component. (Bottom panel) The fit residuals in terms of sigmas with error bars of size 1$\mathrm{\sigma}$.}
 	\label{fig.ic694spec}
 \end{figure}

Therefore we extracted the spectrum of the nucleus of IC 694 (Src 18) from the same annulus and using the same background as in the photometric analysis (\S \ref{sourcedetection}; Table \ref{tab.properties}). Src 18 was strongly contaminated from the local background (see Fig. \ref{fig.ic694spec}) in the soft band. In order to assess the effect of the background in our analysis we also tried to model the spectrum using a background region outside the D25 area of the galaxy. As expected the errors in the soft ($\mathrm{<1.5\ keV}$) part of the spectrum were significantly smaller. Apart from that no significant change in the shape of the spectrum was noticed. Therefore we decided to use the local background of the source.

We obtained a good fit with an absorbed power-law model ($\mathrm{\chi^2_{\nu}/dof=40.10/31}$). The best-fit spectral parameters for all models are shown in Table \ref{tab.ic694}. However, because there is a hint for an emission line at $\mathrm{\sim6.0-7.0\ keV}$ we introduced a Gaussian line in our model (model in XSPEC: \texttt{phabs(po+gaussian)}).

If we model it as an unresolved line its centre is at $\mathrm{6.60_{-0.05}^{+0.10}\ keV}$  which is consistent to an ionized Fe-K$\alpha$ line. The inclusion of the Gaussian line improves the fit, but at a non-statistically significant level, and it makes the photon index slightly steeper ($\sim 1.50$) with respect to $\Gamma\sim0.93$. As a result the absorption corrected flux in the broad and soft bands change by 30\% and 140\% respectively (the hard flux is effectively the same). These results still do not give any evidence for either a weak, relatively obscured AGN, or a buried AGN.  Either case is consistent with the non-detection of a $\mathrm{6.4\ keV}$ Fe-K$\mathrm{\alpha}$ line. On the other hand, a $\mathrm{6.7\ keV}$ line from highly ionized iron could be produced by a high temperature thermal plasma associated with intense star-forming activity confined in the nuclear region of 	IC 694.

\begin{table*}
	\centering
	\begin{minipage}{140mm}
		\caption{Nucleus of IC 694 spectral fitting parameters}
	\begin{tabular}{@{}ccccccccc@{}}
			\hline    			
			model &$\mathrm{N_H}$ & $\Gamma$ & Norm\footnote{Normalization of power-law in units of $\mathrm{10^{-5}\ photons\ keV^{-1}\ cm^{-2}\ s^{-1}}$ at 1 keV.} & E & Norm\footnote{Normalization of Gaussian line in units of $\mathrm{10^{-6}\ photons\ keV^{-1}\ cm^{-2}\ s^{-1}}$.}  & $\mathrm{\sigma}$ & EW & \\
        &$10^{22}\ cm^{-2}$ & & & keV &     &   keV  & keV &   $\chi^2/dof$ \\ 
			\hline
			\hline
        phabs(po+gaussian) &$3.62_{-1.70}^{+2.80}$ &	 $2.10_{-0.93}^{+1.30}$ &	$5.49_{-3.97}^{+32.17}$ & $6.57_{-0.29}^{+5.3}$ & $3.40_{-1.70}^{+18.90}$ &$0.55_{-0.23}^{+2.95}$ & $3.11$& 28.47/28 \\[5pt]        
        phabs(po+gaussian)&$2.72_{-1.32}^{+2.00}$ &	 $1.50_{-0.74}^{+0.88}$ &	$2.37_{-1.62}^{+6.98}$ & $6.60_{-0.05}^{+0.10}$ & $1.64\pm 1.02$ & 0.02\footnote{This parameter is frozen.} & $1.19$& 33.01/29 \\[5pt] 
        phabs\,po&$1.87_{-1.00}^{+1.47}$ &	 $0.93_{-0.60}^{+0.70}$ &	$1.02_{-0.62}^{+2.02}$ & -&- & -& -& 40.10/31 \\
         			\hline
		\end{tabular} 	
		\label{tab.ic694}
	\end{minipage}
\end{table*}

\section{DISCUSSION}\label{discussion}

In the previous sections we presented the results from the analysis of the discrete sources (photometry, spectral, timing analysis), as well as the spectral properties of the integrated galactic emission. In this section we discuss these results in the context of the star-forming of the star-forming activity in Arp\,299. We examine the contribution of the XRBs, ULXs, and the diffuse emission to the integrated luminosity of the Arp\,299. We discuss the nature of the discrete sources and the implication of our results for studies of very high SFR galaxies. 

\subsection{Multiwavelength comparison}\label{multiwavelengthcmparison} 

To further explore the nature of the discrete sources we used Fig. \ref{fig.color} where we combined  HST images in the F814W ($\mathrm{814\ nm}$; green) and F435W ($\mathrm{435\ nm}$; blue) and IRAC band 4 ($\mathrm{8\ \mu m}$; red) non-stellar image. We see that the majority of the sources are located in the most actively star-forming regions with very few sources located outside the main body of the galaxy, indicating that they are HMXBs.

We can further explore this assertion by looking at the ages of the star-forming episodes in the regions of our sources. The ages of the nucleus of IC 694 (A) and the overlap region (C-C'), based on the ratio of mid infrared lines, are $\mathrm{>7\ Myr}$ and $\mathrm{4-7\ Myr}$ respectively, with complex C-C' considered to be the youngest region of Arp\,299 \citep{ah09}. This association indicates that Src 18 (Nucleus of IC 694; A), Src 7 (nucleus of NGC 3690; B1), Src 4 (C), and Src 10 (C") are young HMXBs; region C' is not associated with any discrete X-ray source (Fig. \ref{fig.color}). Complex C is identified as Src 4 in our sample which is a ULX with $\mathrm{L(0.1-10.0\ keV)= 1.03\times 10^{40}\ erg\ s^{-1}}$.  Since this is the youngest region in the galaxy, and ULXs are related to young stellar populations, such a high luminosity would be justified by a very young HMXB. Such a young X-ray binary would be expected to have a very massive black hole fed by a donor star with a strong stellar wind especially at the relatively high metalicity of Arp\,299 giving rise to high X-ray luminosity.

In addition we constructed a map of specific star formation rate (sSFR; defined as the SFR per unit stellar mass) of Arp\,299. We used images in the $\mathrm{8\ \mu m}$ and $\mathrm{3.6\ \mu m}$ obtained from the \textit{Spitzer} IRAC camera provided by \citet{br}. We transformed the $\mathrm{8\ \mu m}$ non-stellar image to SFR map using the relation :
\begin{equation}
\mathrm{SFR_{8\mu m}(M_{\odot}\ yr^{-1})=\frac{\nu L_{\nu}[8 \mu m]}{1.57\times 10^9 L_{\odot}}}
\end{equation}
 \citep{wu} and the $\mathrm{3.6\ \mu m}$ image to a mass map using the relation:
\begin{equation}
\mathrm{\log \frac{M}{M_{\odot}}=(-0.79\pm 0.03)+(1.19\pm 0.01)\times \log \frac{L(3.6\mu m)}{L_{\odot}}}
\end{equation}
\citep{zhu10}. We therefore computed that the total D25 sSFR of the galaxy  is $\mathrm{sSFR=1.54\times 10^{-10} yr^{-1}}$.
\citet{leh10} suggest a limit in sSFR of $5.9\times 10^{-11}\ yr^{-1}$, above which the X-ray source population are dominated by HMXBs. Since the integrated sSFR of Arp\,299 as well as the sSFR at the location of the X-ray sources are above this limit, we can conclude that all the sources we have detected are HMXBs. This is in good agreement with the maximum number of LMXBs expected in our sample, less than 2 sources, based on the XLF for LMXBs of \citet{leh14}, and stellar mass calculated from the K-band luminosity of the galaxy. 

Furthermore, we wanted to see if a correlation between sSFR and the youngest most massive stars exists in Arp\,299. In this case we could use this correlation to identify which sources are younger and possibly hosting more massive donor stars by inspecting the sSFR map of the galaxy. We used a map of [NeIII]/[NeII], which is a tracer of the most massive stars present in a galaxy \citep{ah09}.
Comparison of this map with the sSFR map of the galaxy did not show any correlation between the [NeIII]/[NeII] ratio and the sSFR at the general region of the X-ray sources. The fact that we do not see a correlation most probably implies that the highest values of the [NeIII]/[NeII] ratio could correspond to deeply embedded areas of the galaxy, since these mid-infrared lines are less affected by absorption, than the shorter wavelengths used to trace the SFR and stellar mass.

\begin{figure}
	\resizebox{\hsize}{!}{\includegraphics[scale=1]{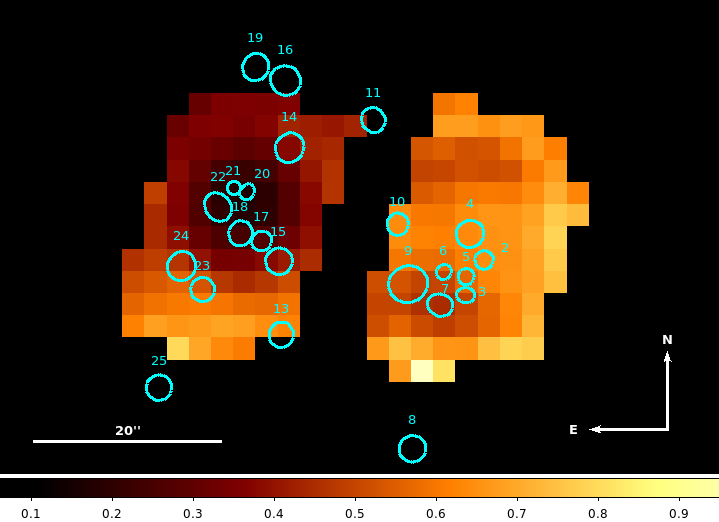}}
	\caption{A [NeIII] $\mathrm{15.56\ \mu m}$/[NeII] $\mathrm{12.8\ \mu m}$  line ratio map of Arp\,299 \citep{ah09}. The X-ray sources are overlaid with cyan ellipses. Lighter colours indicate higher [NeIII]/[NeII] ratios indicating younger star-forming regions (see text for details).}
	\label{fig.herrero}
\end{figure}

\subsection{The Star-formation rate}\label{thestarformationrate}

In order to study the link between X-ray binaries, the hot gas of the galaxy, and its star-forming activity we calculate the total star formation rate (SFR) of Arp\,299 by adopting two different methods. Based on \textit{IRAS} flux densities and using the relation of \citet{helou} we find that the total infrared luminosity is $\mathrm{L_{(8-1000\ \mu m)}=5.9\times 10^{11}\ L_{\odot}}$ and then using the calibration of \citet{ken12} we find a $\mathrm{SFR_{IR}=88.89\ M_{\odot}\ yr^{-1}}$. Since the total infrared luminosity includes the contribution of the AGN, the SFR estimated this way should be treated as an upper limit to the total SFR of the galaxy.

We have also calculated the SFR from the polycyclic aromatic hydrocarbon (PAH) emission, which is a reasonably good SFR indicator \citep{wu, shipley}, using an IRAC $\mathrm{8\mu m}$ non stellar image \citep{br} and the calibration relation of \citet{wu}. This gives $\mathrm{SFR_{8\mu m}=33.06\ M_{\odot}\ yr^{-1}}$. The image was saturated at the nuclei and knowing that the $\mathrm{8\mu m}$ PAHs emission is suppressed in AGN \citep{lutz} we consider this as lower limit.

\subsection{Nature of the X-ray sources}\label{natureofthexraysources}

Based on analysis presented in section \ref{luminosityofxraybinariesandulxs} the total luminosity ($\mathrm{0.1-10\ keV}$) of the X-ray binaries based on their power-law component is $\mathrm{2.5\times 10^{40}\ erg\ s^{-1}}$ (absorbed) and $\mathrm{48.0\times10^{40}\ erg\ s^{-1}}$ (unabsorbed). From the results in Table \ref{tab.binulxslum} we find that the contribution of binaries to the absorbed luminosity of Arp\,299 is 52\% in the broad band, 21\% in the soft band, and 77\% in the hard band, while their contribution to the unabsorbed luminosity is 39\% in the broad band, 28\% in the soft band and 77\% in the hard band.

We can also compare the number of X-ray sources using their scaling relation $\mathrm{N_{XRB}(>10^{38})=3.22\times SFR}$ \citep{m1}. Although our observation does not reach luminosities all the way down to $\mathrm{10^{38}\ erg\ s^{-1}}$ due to incompleteness, we can still have an idea of the number of XRBs. According to the integrated SFR of the galaxy (\S \ref{thestarformationrate}) we would expect to observe between 4 to 11 times more XRBs than we do, depending on the SFR adopted.  

\begin{figure}
 		\resizebox{\hsize}{!}{\includegraphics[scale=1]{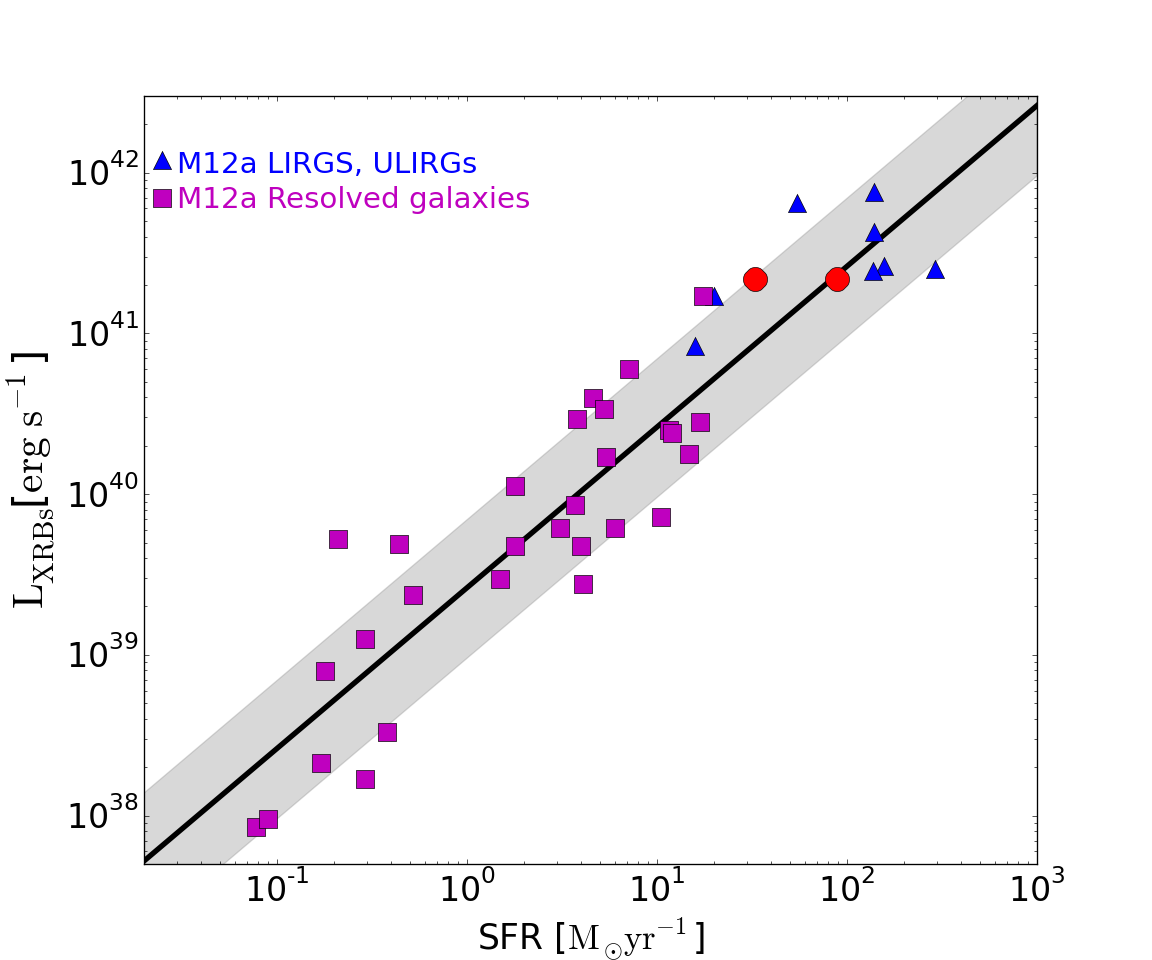}}
 	 	\caption{$L_X-SFR$ relation for a sample of 29 nearby galaxies LIRGS and ULIRGS, the best-fit solid line and the 1$\sigma$ scatter indicated by the grey area  \citep[][M12a]{m1}. We show the position of Arp\,299 on the diagram with two red circles corresponding to the higher and lower SFR we calculated. For Arp\,299 we show the Lx of the power-law which traces the XRBs.}
 	\label{fig.mineolxrbs_sfr}
\end{figure}

Given the much higher SFR of Arp\,299 compared to the population of nearby galaxies we wish to test whether Arp\,299 verifies the scaling relations between XRB X-ray luminosities and SFR observed for other galaxies. For that reason we used the linear relation $\mathrm{L^{XRBS}_{0.5-8.0keV}(ergs^{-1})=2.61\times 10^{39} SFR (M_{\odot} yr^{-1})}$ \citep{m1} between the integrated luminosity of HMXBs and SFR which is based on a sample of 29 nearby star-forming galaxies. The scatter of this relation is 0.43 dex. Measuring the total luminosity of the XRBs of Arp\,299 (Table \ref{tab.binulxslum}) in the same energy range we find $\mathrm{L_{XRBS(0.5-8.0keV)}=2.17\times 10^{41}\ erg\ s^{-1}}$. In Fig. \ref{fig.mineolxrbs_sfr} we plot the X-ray luminosity of XRBs in the sample of galaxies used by \citet{m1} against their SFR, along with their best-fit linear relation. In the same plot we show Arp\,299 (red points depending on the SFR indicator used). We see that Arp\,299 verifies the relation within the 1 $\sigma$ scatter using both of the calculated SFRs.  

We also calculated the expected luminosity of the ULXs again in the ($\mathrm{0.5-8.0\ keV}$) band from the XLF of \citet{m1} and found a luminosity of $\mathrm{L_{ULXs}=(2.3\pm 0.4)\times 10^{41}\ erg\ s^{-1}}$ based on the total infrared SFR, and $\mathrm{L_{ULXs}=(8.5\pm 0.2)\times 10^{40}\ erg\ s^{-1}}$ based on the PAH $\mathrm{8\mu m}$ SFR. The errors in the luminosities were calculated by propagating the errors from the XLF. We find that the observed integrated luminosities of ULXs in the $\mathrm{0.5-8.0\ keV}$ band agree with those expected from the general HMXB XLF within 2.5$\mathrm{\sigma}$ if we use the $\mathrm{SFR_{IR}=88.89\ M_{\odot}\ yr^{-1}}$ and under 1$\mathrm{\sigma}$ if we use the $\mathrm{SFR_{8\mu m}=33.06\ M_{\odot}\ yr^{-1}}$.

From the same general HMXB XLF we calculated that the expected number of ULXs per unit of SFR is  $N_{ULX}^{XLF}/SFR\ (M_{\odot}\ yr^{-1})=0.66$. In the case of Arp\,299, given the total population of 20 ULXs we find a much lower ratio of 0.23 assuming the IR-based SFR, and a ratio of 0.60 assuming the  $\mathrm{8\ \mu m}$ PHAs-based SFR. The same results for the 14 point-like ULXs are 0.16 and 0.42 for the IR-SFR and the PHA-SFR respectively. 

If we now compare with the number of ULXs expected from the work of \citet{swartz11} (($N_{ULX}^{XLF}/SFR\ (M_{\odot}\ yr^{-1})=2.0$)) the deficit is even larger since the values for Arp\,299 are 0.16 to 0.60 (depending on SFR and number of ULXs chosen). We would expect 65-178 ULXs in our observation depending on the SFR chosen.

These results indicate a marginally statistically significant deficit in the number of ULXs in Arp 299:  0.5-1.3$\sigma$ comparing with the XLF of \citet{m1}, and $>2.5-3\sigma$ with respect to with the ULX SFR calibration  of \citet{swartz11} (($N_{ULX}^{XLF}/SFR\ (M_{\odot}\ yr^{-1})=2.0$)). Since this effect in Arp299 (but also other LIRGs)  has been discussed extensively in the literature, we discuss its origin in the light of the deeper observations presented in this work.  

The same deficit is reported by \citet{luan} who studied a sample of 17 nearby LIRGs (including Arp\,299). Although they found 8-9 ULXs for Arp299, based on the first much shallower \textit{Chandra} exposure and the exclusion of sources very close to the nuclei, the deficit remains
when we include the new detections. This result is also supported by \citet{smith} where the total population number of ULXs in LIRGs (including Arp\,299) compared to the far-infrared luminosity is deficient in comparison to the values found in spiral galaxies. \citet{luan} discuss that metallicity may have some influence on ULX numbers \citep{linden,mapelli,bz13a,bz13b,prest,brorby} but they argue that the main reason for this deficit is high columns of gas and dust that obscure these ULXs from our view. 
 
But can high columns of dust and gas be the main reason for this deficit of ULXs in Arp\,299? The X-ray luminosity of HMXBs we observe agrees with the one we would expect and follows the existing scaling relation by \citet{m1} (Fig. \ref{fig.mineolxrbs_sfr}) which strongly suggests that we are not missing any X-ray  emission from sources in heavily obscured areas of the galaxy. If high obscuration was present we would also expect a deficit in the observed X-ray luminosity, or in other words a lower ratio between the observed ULX X-ray luminosity and that expected from the scaling relation with SFR, since the SFR is based on the $\mathrm{8\mu m}$ luminosity which is less affected by the extinction. Therefore, Arp\,299 would be below the line indicating the \citet{m1} relation in Fig. \ref{fig.mineolxrbs_sfr}, which is not the case as it falls on the fitting line for the IR-SFR and above it for the $\mathrm{8\ \mu m}$ SFR. Checking area A (Src 18), one of the areas with the greatest absorption \citep[$\mathrm{A_v=34\ mag}$;][]{ah09}, and using the relation $\mathrm{N_H=1.9\times 10^{21} A_v}$ \citep{zombeck} we find an expected $\mathrm{H_I}$ column density of $\mathrm{N_H=6.46 \times 10^{22}\ cm^{-2}}$ which agrees within the errors with our observed value $\mathrm{3.62_{-1.70}^{+2.80}\times 10^{22}\ cm^{-2}}$. This indicates that the region where most of the HMXBs are expected to be produced is hidden only behind moderate absorbing columns, which would influence the detectability of only the faintest sources, and certainly not their integrated emission. 

Can metallicity be a reason of this deficit?
Recent studies have shown that the numbers of ULXs could be suppressed in high metallicity environments \citep{linden,mapelli,bz13a,bz13b,prest,brorby}. \citet{luan} based on the results of \citet{prest} argued that metallicity on its own could not be a sufficient reason for this deficit. According to \citet{douna} HMXBs are typically 10 times more numerous per unit of SFR in low-metalicity galaxies ($\mathrm{12+log(O/H)< 8.0}$). In Arp\,299 where $\mathrm{12+log(O/H)=8.80}$ \citep{relano} the metallicity could play a role in the deficit of sources but the existing trends \citep{douna} have significant scatter and therefore do not allow us to draw any definite conclusions.
 
In Arp\,299 there is a clear deficit in the number of HMXBs as well as in the number of ULXs but no deficiency in their total observed luminosity compared to that expected from scaling relations. We argue that this is the result of source confusion owing to the large distance of the galaxy (44 Mpc). At this distance the typical scale of star-forming region ($\mathrm{\lesssim 0.5\ kpc}$) corresponds to 2.5\arcsec\ hindering the detection of individual discrete sources especially in the most active star-forming regions. HST data show that star-forming regions have angular scales of $0.5-1.0''$ which are easily confused with the $0.5''$ \textit{Chandra} beam. 
Therefore, given that the total luminosity of the HMXBs is consistent with that expected from the scaling relation, and accounting for the large distance of Arp\,299, we attribute this deficit mainly to confusion effects.
Such effects may also explain the deficit of ULXs in other (U)LIRGs, which are generally in larger distances than Arp\,299.

\subsection{Diffuse emission}\label{diffuseemmision}

Based on analysis presented in section \ref{luminosityofthediffuseemission} we found that the luminosity of the diffuse emission in the soft ($\mathrm{0.1-2.0\ keV}$) band is $\mathrm{1.5\times 10^{41}}$ (absorbed), $\mathrm{1.7\times 10^{41}}$ (unabsorbed), and in the hard ($\mathrm{2.0-10.0\ keV}$) band is $\mathrm{7.1\times 10^{40}}$ (absorbed), and $\mathrm{7.2\times 10^{40}}$ (unabsorbed). From the results presented in Table \ref{tab.binulxslum} we see that the contribution of the soft, diffuse, emission to the total soft X-ray luminosity of the galaxy is 70\% (absorbed), 17\% (corrected for the Galactic absorption), and 18\% (totally unabsorbed). 
The contribution of the hard diffuse emission to the total hard luminosity of the galaxy is 26\% (absorbed) and 25\% (corrected for the Galactic absorption and the totally unabsorbed).

The parameters of the thermal emission such as the temperature ($\mathrm{kT=0.72\pm 0.03\ keV}$; Table \ref{tab.specparam_ulxs}) are consistent within the errors with results found in other star-forming galaxies \citep[e.g. Antennae, M101][]{fabbiano,baldii,kuntz,m2}. The elemental abundances are generally expected to be solar or subsolar in starburst galaxies or ULIRGs \citep[e.g.][]{huo04,strickland02,strickland04}. The abundance of Fe is consistent ($\mathrm{0.26_{-0.03}^{+0.04}}$) with this result. The abundances of Ne and Mg are super solar ($\mathrm{1.24_{-0.29}^{+0.34}}$ and $\mathrm{1.14_{-0.20}^{+0.25}}$) but consistent with solar within the uncertainties.
Subsolar values of Fe and especially the high values of [Mg/Fe] and [Ne/Fe] ratios indicate enrichment of the ISM by Type II supernovae \citep{baldiii}. This is expected given its strong recent star forming activity which would result in an enhanced type II supernova rate.

\begin{figure}
 	\resizebox{\hsize}{!}{\includegraphics[scale=1]{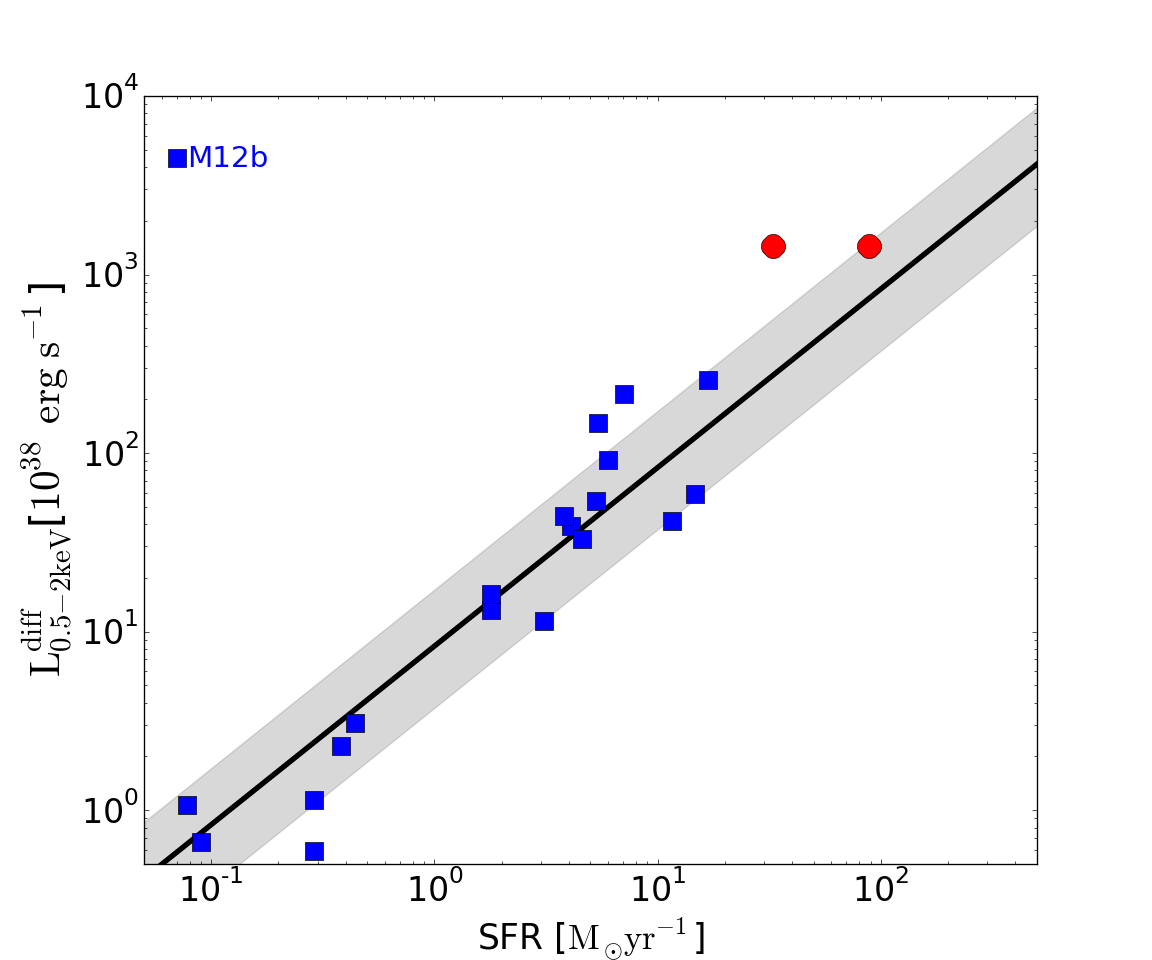}}
 	\caption{Unabsorbed (corrected for the galactic column density) luminosity of the diffuse emission in the $\mathrm{0.5-2.0\ keV}$ band versus SFR for a sample of 21 galaxies, the best-fit solid line, and the 1$\sigma$ scatter indicated by the grey area  \citep[][M12b]{m2}. We show the position of Arp\,299 with the red circles for the two SFRs calculated.}
 	\label{fig.mineoldiff_sfr}
 \end{figure}

\begin{figure}
 	\resizebox{\hsize}{!}{\includegraphics[scale=1]{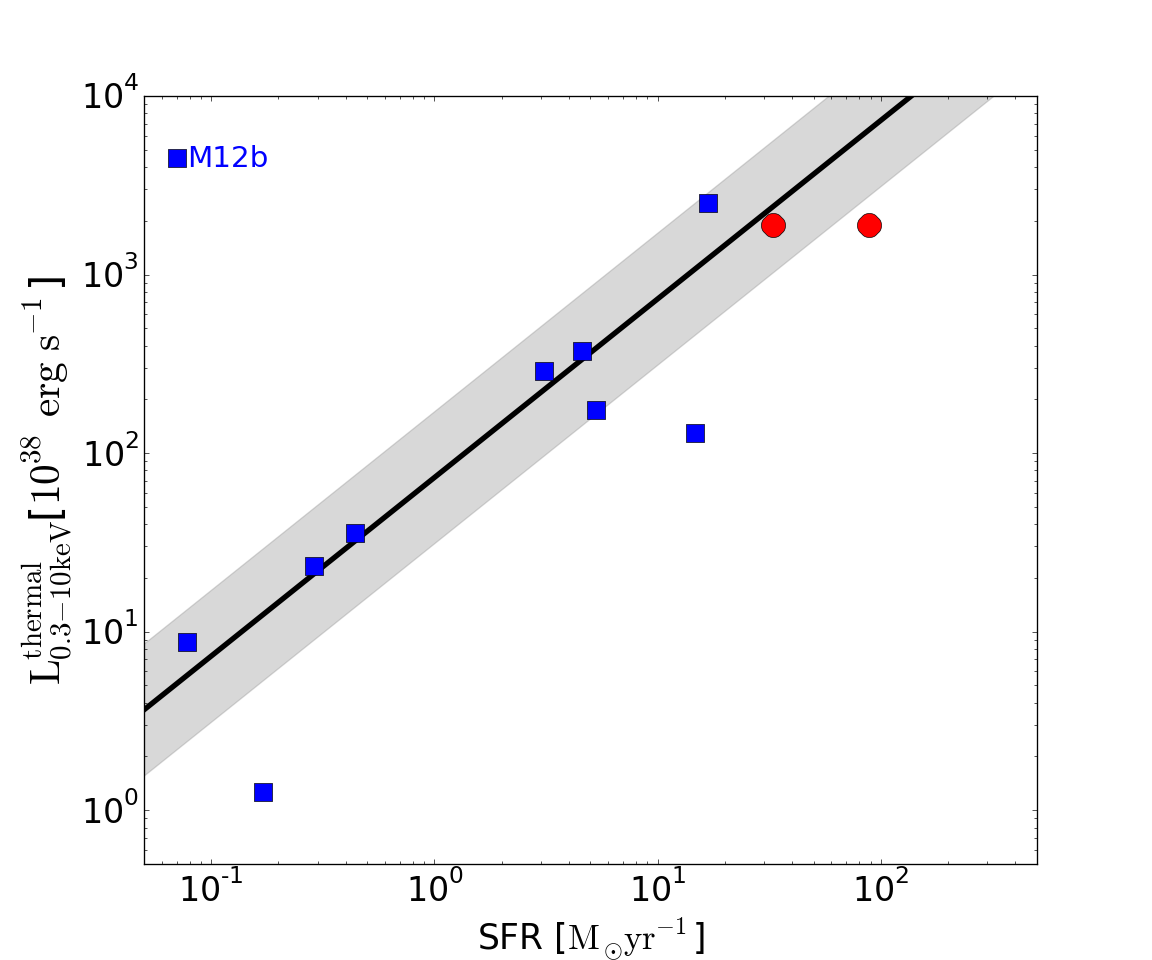}}
 	\caption{Unabsorbed luminosity of the thermal component of the diffuse emission in the $\mathrm{0.3-10.0\ keV}$ band versus SFR for a sample of 21 galaxies and the best-fit solid line, and the 1$\sigma$ scatter indicated by grey area \citep[][M12b]{m2}. We show the position of Arp\,299 with the red circles for the two SFRs calculated.}
 	\label{fig.mineolapec_sfr}
 \end{figure}

\citet{m2} have reported scaling relations of the total diffuse emission from star-forming galaxies and their SFR. Arp\,299 is an extreme star-forming galaxy with a high star-formation rate similar to that witnessed in higher redshift galaxies. Therefore it is instructive to see if it follows this general scaling relation derived from more moderate systems. We used the correlations of \citet{m2} involving (a): the luminosity of the diffuse emission in the $\mathrm{0.5-2.0\ keV}$ band versus SFR, and (b): the luminosity of the thermal component in the $\mathrm{0.3-10.0\ keV}$ band.
The measured luminosities for the diffuse and thermal plasma components of Arp\,299 are $\mathrm{L^{dif}_{0.5-2.0\ keV}=1.44\times 10^{41}\ erg\ s^{-1}}$ and $\mathrm{L^{thermal}_{0.3-10.0\ keV}=1.89\times 10^{41}\ erg\ s^{-1}}$.
In Fig \ref{fig.mineoldiff_sfr} and \ref{fig.mineolapec_sfr} we plot the scaling relation of \citet{m2} along with the values for Arp\,299.

As we see from these Figures Arp\,299 even though has a SFR 2-4 times higher than the most actively star-forming system used in the scaling relation it does not deviate from the general trend. Such deviation would be expected, for example in the case of strong superwinds which would drive significant amount of hot gas outside the galaxy \citep{strickland}. In fact Arp\,299 has a very high supernova rate \citep[e.g.][]{ulvestad,kankare} which would drive such an outflow. As discussed in \S \ref{luminosityofthediffuseemission} there is evidence of a plume extending beyond the optical body of the galaxy in the west of NGC 3690 (Fig. \ref{fig.superwind}), suggestive of such an outflow.
However, this is not a full-scale superwind as seen in other galaxies \citep[e.g. M82;][]{watson,fabbiano88,strickland97,stevens,strickland09}. This could be the result of either the young age of star-formation in Arp\,299 or its much larger mass compared to that of the dwarf galaxies typically exceeding large scale outflows.

\section{SUMMARY}\label{summary}

We have analysed a deep $\mathrm{90\,ks}$ \textit{Chandra} ACIS-S observation of the interacting star-forming galaxy Arp\,299 (NGC 3690/IC 694) and found in total 25 discrete sources and a diffuse emission component. We have found, in agreement with previous studies \citep{lira, fabbianorev}, that the discrete sources dominate the hard X-ray emission of the galaxy above 2.0 keV whereas the diffuse emission component contributes to the soft emission of the galaxy below 2.0 keV.

We verify the presence of an AGN in the nucleus of NGC 3690 by observing its prominent Fe-K$\alpha$ line. However, its contribution to the overall hard emission of the galaxy is quite small ($\sim 20\%$). For the nucleus of IC 694 we have only a hint of the ionized Fe-K$\alpha$ line which could arise from hot thermal gas in this intensively star-forming region. The lack of a neutral iron line does not let us to conclude the existence of an AGN.

We have reported 23 off-nuclear sources, 20 of which exceed the ULX limit. Fourteen of them are point-like sources and therefore considered ULXs. We argued that these sources are young HMXBs based on their association with regions of active, on-going star-formation.

We reported a deficit in the number of HMXBs and ULXs versus the SFR according to the expecting number from the XLF of \citet{m1} and the calibration of the ULX/SFR of \citet{swartz11}. We attributed the main cause of this deficit to confusion of sources since we observed the expected emission of the HMXBs and ULXs.

These results show that although Arp\,299 is a local extreme star-forming galaxy, its observed hard X-ray luminosity per unit SFR is very similar to that of higher-z galaxies \citep[e.g.][]{antara,m3}. This indicated that Arp\,299 is an excellent analogue of the X-ray luminous higher-z objects detected in the deep X-ray surveys.

\section*{Acknowledgements}
We gratefully thank Almudena Alonso-Herrero for providing the  NeIII/NeII \textit{Spitzer} map in electronic form.
The research leading to these results has received funding from the European Research Council under the European Union's Seventh Framework Programme (FP/2007-2013) / ERC Grant Agreement n. 617001. AZ acknowledges support from NASA \textit{Chandra} grant G03-14124X.
LB and RDC acknowledge support from the Italian Space Agency (contract ASI INAF NuSTAR I/037/12/0).
We also made use of the NASA's Astrophysics Data System and observations made with the NASA/ESA Hubble Space Telescope, and obtained from the Hubble Legacy Archive, which is a collaboration between the Space Telescope Science Institute (STScI/NASA), the Space Telescope European Coordinating Facility (ST-ECF/ESA) and the Canadian Astronomy Data Centre (CADC/NRC/CSA).

\newpage
\appendix

\section{Tables}

\begin{table*}
	\centering
		\begin{minipage}{140mm}
		\caption{Properties of the discrete sources in the hard, medium and soft bands.}
		\begin{tabular}{@{}crccrccrrc@{}}
			\hline   
			Src\footnote{Column 1: The source identification number, Columns 2, 5, and 8: Net source counts and corresponding error counts in the hard ($\mathrm{2.0-7.0\ keV}$), medium ($\mathrm{1.2-2.0\ keV}$), and soft ($\mathrm{0.5-1.2\ keV}$) bands respectively, Columns 3, 4, 6, 7, 9 and 10: The background source counts and the signal to noise ratio of each source and hard, medium and soft bands respectively.}  &  Net counts(hard)  & Bkg & S/N & Net counts(medium)  & Bkg & S/N & Net counts(soft)  & Bkg & S/N \\
			ID&  $\pm$error &             &     &      $\pm$error     &    &        &  $\pm$error   &       &    \\
			(1)&(2)&(3)&(4)&(5)&(6)&(7)&(8)&(9)&(10)\\
			\hline 			 
			1      &	59.8 $\pm$   8.9 &  1.1  & 6.5 &  57.3  $\pm$    8.5  &  1.6      & 6.3   &  18.1 $\pm$ 5.6    &  1.9   &   2.9     \\
			2      &	157.0 $\pm$  14.4 &  15.9  & 10. &  156.2  $\pm$    15.6  &  40.8     & 9.3   &  117.5  $\pm$ 15.1  &  58.5    &   7.0     \\
			3      &	210.4 $\pm$  20.5 &  101.0  & 9.6 &  126.2  $\pm$    18.2  &  100.7    & 6.5   &  -0.5  $\pm$ 10.9  &  58.5   &   -0.0     \\
			4      &	113.0 $\pm$  16.6 &  60.0  & 6.7 &  234.0  $\pm$    24.6  &  153.0    & 9.5   &  236.0  $\pm$ 24.9  &  159.0   &   9.4     \\
			5      &	74.4  $\pm$  11.7 &  21.6  & 6.0 &  98.6   $\pm$    13.4  &  30.4     & 7.0   &  33.2   $\pm$ 10.6  &  32.8    &   2.9     \\
			6      &	87.5 $\pm$   11.8 &  22.5  & 6.7 &  54.5  $\pm$    10.7  &  29.5     & 4.5   &  -1.5 $\pm$ 7.0  &  28.5   &   -0.1     \\
			7      &	398.0 $\pm$  24.0 &  80.0  & 15. &  315.2  $\pm$    23.0  &  103.7    & 13.0   &  179.2  $\pm$ 18.7  &  80.7   &   9.0     \\
			8      &	11.5 $\pm$   4.6 &  0.5 & 2.2 &  16.5  $\pm$    5.2   &  0.5    & 2.9   &  7.2 $\pm$ 4.6  &  4.7   &   1.3     \\
			9      &	5.4 $\pm$    9.6 &  21.5  & 0.6 &  11.2  $\pm$    15.1  &  64.8     & 0.8   &  55.6 $\pm$ 18.4  &  86.4   &   3.3     \\
			10     &	121.9 $\pm$  12.5 &  8.05  & 9.3 &  84.5  $\pm$    10.9  &  9.5      & 7.3   &  14.8 $\pm$ 5.9  &  6.2   &   2.2     \\
			11     &	8.9  $\pm$   4.4 &  1.0  & 1.8 &  10.4  $\pm$    4.7  &  1.6      & 2.0   &  3.8    $\pm$ 4.3  &  4.2     &   0.7     \\
			12     &	65.7 $\pm$   9.4 &  3.3  & 6.6 &  85.2  $\pm$    10.4   &  1.8   & 8.0   &  47.3 $\pm$ 8.6   &  7.7   &   5.0     \\
			13     &	19.9 $\pm$   6.1 &  4.1  & 2.9 &  25.1  $\pm$    6.9   &  6.8     & 3.3   &  25.3 $\pm$ 7.6  &  13.7   &   2.9     \\
			14     &	88.0  $\pm$  10.6 &  2.0   & 8.1 &  74.3  $\pm$    10.3  &  6.7      & 6.9   &  32.7 $\pm$ 9.2  &  21.3   &   3.2     \\
			15     &	50.1 $\pm$   10.8 &  22.8  & 4.4 &  75.2   $\pm$    12.7  &  30.8     & 5.7   &  22.2  $\pm$ 11.6  &  45.8   &   1.8     \\
			16     &	102.9 $\pm$  11.4 &  1.0  & 8.9 &  86.0  $\pm$    10.6  &  1.0      & 8.1   &  24.8 $\pm$ 7.6  &  6.2   &   3.3     \\
			17     &	215.8 $\pm$  17.8 &  41.1  & 11. &  221.1  $\pm$    18.4  &  49.8     & 11.4   &  122.9 $\pm$ 13.5  &  18.1   &   8.8     \\
			18     &	354.3 $\pm$  26.2 &  90.6  & 14. &  77.6  $\pm$    22.0  &  121.3    & 4.0   &  -8.7  $\pm$ 16.6  &  82.7   &   -0.6     \\
			19     &	27.0  $\pm$  6.7 &  2.0   & 3.8 &  20.7  $\pm$    6.3   &  3.3      & 3.1   &  9.0    $\pm$ 6.4 &  12.0    &   1.2     \\
			20     &	6.9 $\pm$    4.1 &  1.1  & 1.5 &  12.1  $\pm$    7.2  &  18.8     & 1.4   &  24.2 $\pm$ 9.8 &  38.8   &   2.1     \\
			21     &	11.7 $\pm$   5.7 &  5.3  & 1.8 &  32.0   $\pm$    8.0  &  10.0     & 3.7    &  1.7  $\pm$ 7.5  &  25.3   &   0.2     \\			 
			22     &	-4.5 $\pm$   8.8 &  31.4  & -0.5&  35.8  $\pm$    14.5  &  73.1     & 2.4    &  100.7  $\pm$ 20.1  &  134.2   &   4.9     \\
			23     &	87.9 $\pm$   11.5 &  16.0  & 7.1 &  38.4  $\pm$    8.9   &  16.6    & 3.9   &  6.7 $\pm$ 6.1  &  13.3   &   0.9     \\
			24     &	33.8 $\pm$   7.8 &  7.1  & 4.0 &  9.9  $\pm$    7.5   &  22.0     & 1.1   &  6.2  $\pm$ 8.3  &  32.8   &   0.6     \\    
			25     &	5.2 $\pm$    3.6 &  0.8  & 1.2 &  11.9  $\pm$    4.7  &  1.0      & 2.3   &  9.4 $\pm$ 5.0  &  5.5   &   1.5     \\
			26     &	14.3 $\pm$   5.0 &  0.6 & 2.6 &  12.6  $\pm$    4.7   &  0.4   & 2.4   &  8.6 $\pm$ 4.1  &  0.4   &   1.9     \\
			\hline
		\end{tabular} 	
		\label{tab.properties2}
	\end{minipage}
\end{table*}

\begin{table*}
	\centering
	\begin{minipage}{140mm}
		\caption{Properties of lower significant detections ($2.0<\mathrm{SNR}<3.0$) in the broad band ($\mathrm{0.5-7.0\ keV}$).}
		\begin{tabular}{@{}cccrrccc@{}}
			\hline   
			Src\footnote{Column 1: The source identification number, Columns 2 and 3: Sky coordinates, Column 4: Net source counts and corresponding error counts, Columns 5 and 6: The background source counts and the signal to noise ratio of each source, Columns 7 and 8: The two ellipse major and minor radius for the source apertures.}  &  RA      & Dec   &       Net counts& Bkg & S/N & $\mathrm{r_1}$  &   $\mathrm{r_2}$  \\
			ID &h m s &$^{\circ}$ $'$ $''$   & $\pm$error & & &$''$&$''$ \\
			(1) &  (2)      & (3)   &       (4)& (5) & (6) &(7)  &   (8)  \\
			\hline
			1a  &  11:28:29.3 & +58:33:50.8  &  24.0 $\pm$ 7.5   & 12.0  &   2.8      &   0.96 & 1.02 \\
			2a  &  11:28:30.4 & +58:33:58.8  &  33.1 $\pm$ 10.5  & 33.9   &   2.9      &   1.45 & 1.45 \\
			3a  &  11:28:31.5 & +58:33:38.1  &  34.4 $\pm$ 11.2   & 42.6   &   2.8      &   1.00 & 0.96  \\
			4a  &  11:28:31.5 & +58:33:55.2  &  21.3 $\pm$ 7.5  & 16.6   &   2.4      &   1.37 & 1.22   \\
			5a  &  11:28:33.1 & +58:33:59.7  &  16.5 $\pm$ 6.1  & 6.5   &   2.4      &   0.95 & 1.22   \\
			6a  &  11:28:33.8 & +58:33:52.5  &  39.5 $\pm$ 13.5 & 63.5   &   2.7      &   0.85 & 0.97  \\
			\hline
			\end{tabular} 	
			\label{tab.properties1a6a}
		\end{minipage}
	\end{table*}

\bsp	
\label{lastpage}
\end{document}